%% file: main.tex
\documentclass[a4paper,12pt]{article}
\usepackage{graphicx}
\usepackage{float}

\newcommand{\be}{\begin{equation}}
\newcommand{\ee}{\end{equation}}
\newcommand{\bea}{\begin{eqnarray}}
\newcommand{\eea}{\end{eqnarray}}

\newcommand{\nn}{\nonumber\\}

\begin{document}

\title{Analysis of flow effects in relativistic heavy-ion collisions
within the CBUU approach
\protect\footnote{Supported by BMBF and GSI Darmstadt}\hspace{3mm}
\footnote{part of the PhD thesis of A. Hombach}}
\author{A. Hombach, W. Cassing, S. Teis and U. Mosel \\
        Institut f\"{u}r Theoretische Physik, Universit\"{a}t Giessen \\
        D-35392 Giessen, Germany}
\maketitle

\begin{abstract}

We study flow phenomena in relativistic heavy-ion collisions,
both in transverse and radial direction, in comparison to experimental data.
The collective dynamics of the nucleus-nucleus collision is described
within a transport model of the coupled channel BUU type (CBUU).
This recently developed version includes all nucleonic 
resonances up to 1.95 GeV in mass and mean-field potentials 
both of the Skyrme and momentum dependent MDYI type.
We find that heavy resonances play an important role
in the description of transverse flow above \mbox{1 AGeV} incident energy.
For radial flow we analyse reaction times and equilibration  
and extract the parameters $T$ and $\beta$ for
temperature and collective flow velocity within different prescriptions.
%which gives rise to substantial differences.
Furthermore, we apply a coalescence model for fragment production
and check the mass dependence of the flow signals.
\end{abstract}
\vspace{2cm}
\noindent
PACS: 25.75. +r                \\
Keywords: Relativistic Heavy-Ion Collisions

\newpage
%===============================================================

\begin{section}{Introduction}

Relativistic heavy-ion collisions (HIC) provide a unique tool to study
nuclear matter at high densities and temperatures. 
However, throughout a collision the system is partly far away from 
equilibrium and both particle production and collective motion
depend on various quantities, such as the stiffness of the
equation-of-state (EoS), the momentum dependence of the interaction or
mean-field potentials (MDI), in-medium modifications of
the $NN$ cross section $\sigma_{NN}$ 
and even the initial momentum distribution of the nucleons
%\cite{GBdG87,WPKdG88,GWPLdG90,ZdGG94,roy97,PD93, bass98,Hartnack98,Ramilien95,LBK90,BKL91}.
\cite{GBdG87}-\cite{BKL91}.
Since these dependencies are in general not very strong or unique
and also model parameters influence the results,
it is necessary not to focus on a single observable alone,
but to investigate the dynamical evolution of the HIC
within a single model that is able to describe all relevant quantities.

In this work we investigate the collective behavior of 
nuclear matter in a heavy-ion collision in the energy range from 
150 AMeV to 2 AGeV for various systems using the CBUU model which 
is briefly presented in Section \ref{BUU}. This model
was shown to describe pion production \cite{teis96},
photoproduction and -absorption \cite{effe1,effe2} as well as 
pion-induced reactions \cite{weid97}.
Recently we investigated isospin equilibration and its dependence
on the EoS as well as on medium modifications of the $NN$ cross sections 
\cite{hom98_1} in comparison to experiments currently performed
at GSI \cite{Yvonne98}.

The paper is organized as follows: In Section \ref{BUU}
we describe the CBUU model used for the calculations
where we especially focus on mean field potentials.
Section \ref{trans_flow} presents the results and discussions
on transverse flow, while Section \ref{rad_flow} is assigned to
the results on radial flow. 
The mass dependence of the flow quantities is investigated in 
Section \ref{mass}, whereas a summary follows in Section \ref{summary}.

\end{section}

%===============================================================
\input{buu}
\input{transfl}
\input{radfl}
\input{mass}
\input{summry}

\input{biblio}
\input{captions}
\input{figures}
%===============================================================
\end{document}

%% file: buu.tex
\begin{section}{The CBUU-Model}
\label{BUU}

%--------------------------------------------------
\begin{subsection}{Basic equations and collision term}

For our study we use the CBUU transport model which has already been
introduced in an earlier publication \cite{teis96}.
Whereas in \cite{teis96} we have concentrated on particle production
and therefore on the collision term, we will focus here on the
mean-field potential, which governs the bulk properties of the 
nuclear matter throughout a heavy-ion collision.

The basic equation of a BUU transport model reads 
\cite{bertsch88,cassing88,cassing90,kweber93,cassing99}:
\bea 
\label{buueq}
& & \frac{\partial f_1}{\partial t} + 
\left\{ \frac{\vec{p_1}}{E_1} + 
\vec{\nabla}_p\, U_1(\vec{r},\, 
\vec{p_1}) \right\} 
\, \vec{\nabla}_r f_1 
- 
\vec{\nabla}_r U_1(\vec{r},\, \vec{p_1}) 
\vec{\nabla}_p f_1 \nn
& & \hspace{1cm}  =  \sum_{2,3,4}
\frac{g}{(2\pi)^3} \, 
\int\! d^3 p_2 \, \int\! d^3p_3 \, \int\! d \Omega_4 \,\,\, 
\delta^3 \left(\vec{p}_1 + \vec{p}_2 
- \vec{p}_3 - \vec{p}_4 \right) \nn
& &\hspace{1.4cm} \times\,\,(
v_{34}\, \frac{d\sigma_{34 \rightarrow 12}}{d\Omega}\, 
f_3\, f_4\, \bar{f}_1\,\bar{f}_2 -
v_{12}\, \frac{d\sigma_{12 \rightarrow 34}}{d\Omega}\, 
f_1\, f_2\, \bar{f}_3\,\bar{f}_4)\quad,\hspace{1.4cm}\mbox{}
\eea
where $f_i$ stands for $f_i(\vec r, \vec p_i, t)$ denoting the
single-particle phase-space distribution function for the
nucleons;
$\bar{f}_i=(1-f_i)$ are the Pauli-blocking factors for fermions.
Since we go beyond the pion production
threshold similar equations have to be solved for the 
nucleon
%$\Delta(1232)$, $N(1440)$, $N(1535)$, the 12 higher 
resonances and for the mesons included.
These transport equations are coupled via the collision term. 
We, therefore, refer to the model as the coupled channel BUU model
(CBUU). Note, that in Eq.~(\ref{buueq}) all phase-space distributions $f$ 
appear with equal space-time argument, since the collision term is 
assumed to be local in space and time.

As already described in \cite{teis96} we include 14
nucleon resonances up to masses of 1.95 GeV/c$^2$,
i.e. the $\Delta(1232)$,
$N(1440)$, $N(1520)$, $N(1535)$, $\Delta(1600)$, 
$\Delta(1620)$, $N(1650)$, $\Delta(1675)$, $N(1680)$, $\Delta(1700)$,
$N(1720)$, $\Delta(1905)$, $\Delta(1910)$ and $\Delta(1950)$, 
where the resonance properties are adopted from the PDG \cite{pdg}.
The mesons incorporated are $\pi$, $\eta$, $\rho$ and $\sigma$, 
where the $\sigma$-meson is introduced to describe correlated 
pion-pairs with total spin $J = 0$. 
For the baryons as well as for the mesons all isospin degrees 
of freedom are treated explicitly.

The r.h.s. of Eq.~(\ref{buueq}), i.e. the collision integral, 
describes the  changes of $f_i(\vec{r}, \vec{p_i}, t)$ due to 
two-body collisions among  the hadrons ($h$):
$h_1\!+\!h_2 \leftrightarrow h_3\!+\!h_4$ 
and two-body decays of baryonic and mesonic resonances ($R,r$) to
hadrons and mesons ($m$):
$R_h \leftrightarrow h\!+\!m, \,\, r_m \leftrightarrow m_1\!+\!m_2$. 
Three body final states are treated as two subsequent 2-body 
processes. The in-medium collision rate is represented by 
%$v_{12}\frac{d\sigma_{12 \rightarrow 34}}{d\Omega}$ 
$v_{12}\cdot d\sigma/d\Omega$ 
where $d\sigma/d\Omega$ is the free differential cross section 
and $v_{12}$ is the relative velocity between the colliding hadrons  
$h_1$ and $h_2$ in their center-of-mass system.
Taking elastic collisions only, 
%$v_{12}\frac{d\sigma_{12 \rightarrow 34}}{d\Omega}$ 
$v_{12}\cdot d\sigma_{12 \rightarrow 34}/d\Omega$ 
equals
%$v_{34}\frac{d\sigma_{34 \rightarrow 12}}{d\Omega}$;
$v_{34}\cdot d\sigma_{34 \rightarrow 12}/d\Omega$;
in case of inelastic collisions one has either to use
detailed balance, or, as we do, divide the cross section up 
into (matrix element $\times$ phase space factors) and
use these matrix elements for the determination of the 
backward reaction \cite{teis_dr}.
For the most important nucleonic resonance, the $\Delta(1232)$,
we use a parametrization following the result of an OBE model
calculation by Dimitriev and Sushkov \cite{dimi86}
for mass- and angle-differential cross sections.

In the collision integrals 
describing two-body decays of resonances the product 
(relative velocity $\times$  cross-section $\times \ f_2$) 
has to be replaced 
by the corresponding decay rate and the proper fermion 
blocking factors in the final channel have to be introduced. 
The factor $g$ in Eq.~(\ref{buueq}) stands for the 
spin degeneracy of the particles participating in the collision 
whereas $\sum_{2,3,4}$ stands for the sum over
the isospin degrees of freedom of particles 2, 3  and 4.

We include the following elastic and inelastic baryon-baryon, 
meson-baryon and meson-meson collisions: 
\bea 
N N & \longleftrightarrow & N N \nn
N N & \longleftrightarrow & N R \nn
N R & \longleftrightarrow & N R'\nn
\label{doubledelta}
N N & \longleftrightarrow & \Delta(1232) \Delta(1232) \\
R & \longleftrightarrow & N \pi \nn
\label{zwopi1}
R & \longleftrightarrow & N \pi \pi \nn
\label{zwopi2}
  & \quad = & \Delta(1232) \pi,\, N(1440)\pi, \, N \rho,\, N \sigma \\
N(1535) & \longleftrightarrow & N \eta \nn
N N & \longleftrightarrow & N N \pi \nn 
\rho & \longleftrightarrow & \pi \pi \, \,\, \, \,  \mbox{(p-wave)}\nn
\sigma & \longleftrightarrow & \pi \pi \, \, \, \, \, \mbox{(s-wave)}, 
\nonumber
\eea
where the $2\pi$-decay of a hadronic resonance, $R$, 
is introduced as subsequent  $1\pi$-decays (\ref{zwopi2})
and the $N N \longleftrightarrow \Delta\Delta$
process (\ref{doubledelta}) is parametrized according to 
Huber and Aichelin \cite{HA94}.

\end{subsection}

%--------------------------------------------------

\begin{subsection}{Mean-field potentials}

The l.h.s. of Eq.~(\ref{buueq}) represents the  
Vlasov-equation for hadrons moving in a momentum-dependent 
field $U(\vec{r}, \vec{p})$, where $\vec{r}$ and $\vec{p}$ 
stand for the spatial and momentum coordinates of the hadrons, 
respectively. 
From Dirac-phenomenological optical-model calculations 
\cite{kweber93,arnold79} it is known that elastic nucleon-nucleus 
scattering data can only be described when using proper 
scalar and vector potentials.
Since this approach has proven to be numerically very difficult
\cite{maru94} we use as starting point the non-relativistic
momentum-dependent mean field potential proposed by Welke et al.
\cite{WPKdG88,GWPLdG90}, i.e.

\be
U(\vec{r}, \, \vec{p}) = 
A \frac{\rho(\vec r)}{\rho_0} + 
B \left(\frac{\rho(\vec r)}{\rho_0}\right)^\tau + 
2 \frac{C}{\rho_0} \int d^3 p'\frac{f(\vec{r},\vec{p'})}
{1 + \left(\frac{\vec{p}-\vec{p'}}{\Lambda}\right)^2} \;.
\label{welkepot}
\ee

This ansatz in principle enables to guarantee energy conservation
since it can be derived from a potential energy density functional.
However, as an extension of the momentum-independent Skyrme type 
potentials for nuclear matter \cite{cassing90,GdG90} 
the parametrization (\ref{welkepot}) has no manifest Lorentz properties,
whereas the latter are required for a transport 
model at relativistic energies. To achieve this goal
we evaluate the non-relativistic mean-field potential $U$ 
for a particle in the local rest frame (LRF) of the surrounding 
nuclear matter which is defined by the frame of reference with 
vanishing local vector baryon current, $\vec{j}(\vec{r},t) = 0$.
For this the reaction volume is divided into a grid with
1 fm$^3$ cell volume and the baryon current is evaluated in each
cell.
In the LRF we then transform the non-relativistic potential 
$U(\vec{r},\vec{p})$ to a scalar one, $U_S$, by  the identification
\be
\sqrt{\vec p^{\,2} + m^2} + U(\vec{r},\,\vec{p}) =  
\sqrt{\vec p^{\,2} + \left(
m + U_S( \vec{r},\,\vec{p}) \right)^2},
\label{defuscal}
\ee
thus defining a local momentum-dependent 
effective mass $m^*=m+U_S$ for the particle.
This effective mass is used 
throughout our calculations for the baryons. 
The mesons are propagated as free particles; 
their effective mass is equal to their restmass, 
i.e. $U_{\rm meson}\equiv 0$.

Due to the relativistic dispersion relation (\ref{defuscal}),
the potential $U_S(\vec{r},\,\vec{p})$  has now definite 
Lorentz-properties.
This enables us to guarantee energy conservation in each two-body 
collision ( $N_1+N_2 \rightarrow N_3+N_4$) as 
\be
\sqrt{\vec p_1^{\,2} + m_1^{*\, 2}} +
\sqrt{\vec{p}_2^{\,2} + m_2^{*\, 2}}
 =  \sqrt{\vec{p}_3^{\,2} + m_3^{*\, 2}} + 
\sqrt{\vec{p}_4^{\,2} + m_4^{*\, 2}}
\label{defencons}
\ee
as well as in resonance decays.

The parameters of the Potential $U$ we fit to match the requirements
\bea
\label{potprop}
&&\left.\frac{E}{A}\right|_{\rho_0}\!\! = -16\;{\rm MeV},\quad
\left.\frac{\partial E}{\partial \rho}\right|_{\rho_0}\!\! = 0, \quad
K = 210\,|\,260\,|\,380 \;{\rm MeV},\quad \nn
&&U(E=300\,{\rm MeV})=0\quad
\mbox{and}\quad U(p=\infty)=+32\;{\rm MeV}\quad.
\eea
These constraints we derive from the results of a microscopic 
calculation from Wiringa et al. \cite{wir88_1,wir88_2}
with Hamiltonians that describe $NN$ scattering data, 
few body binding energies and nuclear matter saturation properties.
In \cite{wir88_2} the results of the calculation based on the 
Urbana v14 model plus additional three-body interaction (UV14+TNI)
is fitted by the approximation 
\be
\label{wirpot}
U=\alpha(\rho) \,+\, 
\frac{\beta(\rho)}{1+\left(\frac{p}{\Lambda}\right)^2}
\ee
for different densities, which is close to the functional form 
of Eq.~(\ref{welkepot}). Thus we fit Eq.~(\ref{wirpot}) with 
(\ref{welkepot}) for normal nuclear matter
density which we take to be $\rho_0 \approx$ 0.168 fm$^{-3}$. 
The best agreement to
(\ref{wirpot}) over the whole density regime from 0.1 to 
\mbox{0.5 fm$^{-3}$}
is achieved using a compressibility of \mbox{$K$ $\simeq$ 230 MeV},
however, we fit the three different compressibilities 
denoted in Eq.~(\ref{potprop}), $K$ = 210, 260 and 380 MeV, to allow for
simulations testing different EoS.

\end{subsection}

%--------------------------------------------------

\begin{subsection}{Numerical realization} \label{numerics}

The CBUU-equation~(\ref{buueq}) is solved by means of the 
test-particle method, where the phase-space distribution function 
$f(\vec r, \vec p,t)$ is represented by a sum over 
$\delta$-functions: 
\be
f( \vec{r},\, \vec{p},\, t) =\frac{1}{N}\, \sum_{i=1}^{N \times A}
\delta \left(\vec{r}- \vec{r}_i(t)\right) \times \delta \left( \vec{p}
- \vec{p}_i(t) \right).
\label{tpansatz}
\ee
Here $N$ denotes the number of testparticles per nucleon while $A$ is
the total number of hadrons in the two colliding nuclei. 
Inserting the ansatz (\ref{tpansatz}) into the CBUU-equation
leads to the equations of motion for the testparticles: 
\bea
\frac{d \vec{r}_i(t)}{d t} & = & \frac{\vec{p}}{E} + \frac{m^*}{E}\;
\vec{\nabla}_p U_S(\vec{r}_i,\, \vec{p}_i(t) ) \nn
\frac{d \vec{p}_i(t)}{d t} & = & - \frac{m^*}{E}\;
\vec{\nabla}_r U_S(\vec{r}_i,\, \vec{p}_i(t) ). \label{tpeoms}
\eea
Thus the solution of the CBUU-equation within the testparticle 
method reduces to a set of
equations of motion for classical point particles (\ref{tpeoms}).
For the actual numerical simulation we discretize the time $t$ 
into steps of typical \mbox{0.5 fm/c}
and integrate the equations of motion employing a predictor-corrector
method \cite{stoer80}.

For the evaluation of the mean-field potential we expand the 
local Thomas-Fermi approach used for the initial momentum distribution
for the testparticles.
Then the integral in Eq.~(\ref{welkepot}) can be solved analytically
giving \cite{WPKdG88} 
\bea
\label{welkepottfa}
U(\vec r,\vec p) &=&  
\pi \Lambda \left[
\frac{p_F^2 + \Lambda^2 - p^2}{2p\Lambda}
\ln\frac{\left(p + p_F\right)^2 + \Lambda^2}{\left(p - p_F\right)^2 + \Lambda^2}
+ \frac{2p_F}{\Lambda}\right. \nn
&&\qquad \left. - 2 \left[\arctan\frac{p+p_F}{\Lambda}
- \arctan\frac{p-p_F}{\Lambda} \right] \right],
\eea
which provides a very fast evaluation of the nucleon potential.
Two-body collisions and resonance decays are treated as described in 
Ref. \cite{teis96}.

This model has been proven to adequately describe pion spectra
\cite{teis96}, pion multiplicities \cite{teis_dr}
and Coulomb effects on pion spectra \cite{teis_coulomb}.

\end{subsection}
\begin{subsection}{Energy conservation}
\label{energy}

One of the most important issues to check for any transport model
is the quality of energy conservation since this influences 
the results on both transverse \cite{GdG90} and radial flow
as well as on particle production.
We evaluate the total energy minus the nucleon restmass 
\bea
E &=& \sum_i t_i + \int \! d^3r\, W(\rho, \vec r) \nn
 &=& \sum_i \left(\sqrt{m_i^2+{\vec p_i}^2} - m_i \right) \nn
&&+ \int \! d^3r\, \left(
A \frac{\rho(\vec r)^2}{\rho_0} + 
B \frac{\rho(\vec r)^{(\tau+1)}}{\rho_0^\tau} + 
\frac{C}{\rho_0} \int \! d^3 p \, d^3 p' \, 
\frac{f(\vec{r},\vec{p}) f(\vec{r},\vec{p'})}
{1 + \left(\frac{\vec{p}-\vec{p'}}{\Lambda}\right)^2}
\right) \nn
&=& \frac{1}{N} 
 \sum_{i=1}^{N\times A}
 \left(\sqrt{m_i^2+{\vec p_i}^2} - m_i \right) \nn
&&+ \sum_{x,y,z} \left(
A \frac{\rho^2}{\rho_0} + 
B \frac{\rho^{(\tau+1)}}{\rho_0^\tau} + 
\frac{C}{N\rho_0} \sum_{j,k} \frac{1}
{1 + \left(\frac{\vec{p_i}-\vec{p_j}}{\Lambda}\right)^2}
\right)\;,
\eea
where the kinetic energy sum runs over all $N\times A $
testparticles and the spacial integral 
$\int \! dxdydz \, W$ is taken as a sum over all grid cells
$\sum_{x,y,z}$ and $\sum_{j,k}$ runs over all particles in the 
current cell $(x,y,z)$. The momenta 
$\vec{p_i},\vec{p_j}$ are taken relative to the LRF.

The upper part of Fig.~\ref{ener_t_Au} shows this total energy per nucleon 
in a \mbox{1 AGeV} central $Au+Au$ collision
in the cascade mode (i.e. neglecting the mean field),
the middle part the results for propagating the nucleons in the mean field 
without collisions and the lower part for the 'full' calculation, 
i.e. propagation of the particles in the mean-field plus collisions.

Whereas the collisional part (upper picture) -- which we require to guarantee
energy conservation in the individual collisions on the
10$^{-5}$-level -- gives rise to a steady total energy loss about 2 \%,
the propagation in the momentum-dependent mean field (middle picture)
in total conserves energy but shows some deviation 
in the high density phase between t=7.5 fm/c to t=25 fm/c.
These deviations from the initial total energy per nucleon
are due to a gaussian smearing algorithm for the density 
distributions 
used to obtain smooth potentials in coordinate and momentum space.
This avoids steep gradients
for numerical reasons which would lead to an artificial and unphysical
acceleration of the particles. 
The smearing, of course, smoothes also steep physical density
gradients, e.g. in the region of the fireball surface.
However, taking the 'bare' density distribution for the
evaluation of the potentials leads to numerically 
instable nuclei with the consequence that the results 
depend on initial conditions as extensively discussed 
in Ref.~\cite{Hartnack98};
with smearing they are stable.

The full calculation (lower picture) shows mainly the behaviour of 
the cascade with a total energy loss of about 1.8 \%, which is 
an excellent value for the following investigations.

\end{subsection}
\end{section}

%% file: transfl.tex
\begin{section}{Transverse flow}
\label{trans_flow}

Within the CBUU model described in the previous Section we now calculate
the transverse flow for various systems and beam energies and 
analyse the dependence on different quantities.

In general, in a HIC three stages can be identified: 
i) the initial phase, which is characterized 
by high relative momenta of the incoming nucleons, ii) the
high compression phase, where due to the high density a lot of
collisions happen thus driving the system towards thermal and 
chemical equilibrium, and iii) the expansion phase, where still 
some particle freeze-out happens, but the collision 
rate drops to zero.
The  phases i) and ii) are most important for collective flow
as will be shown in the following.

%-------------------------------
\begin{subsection}{Origin of flow}

In the participant - spectator picture
transverse flow 
has its origin in the deflection of the spectators at the hot
and dense reaction zone, the fireball (see, e.g. \cite{Metag98}). 
However, in a transport model the flow pattern of the nucleons 
in coordinate space (Fig.~\ref{au1b6movie}) for a $Au+Au$ 
collision at 1 AGeV
and impact parameter b=6 fm shows that the direction of motion of
the spectators in the final state of the reaction, 
clearly characterized by a density close to $\rho_0$,
is nearly unchanged. 
At the back of these spectators participating matter escaping from the
fireball streams in 
outward direction alongside the spectator surface, attracted by the 
nuclear potential field.
Since the flow $F$ is defined as the slope of the transverse momentum
distribution at midrapidity
\be
F=\left.\frac{d\langle p_x\rangle}{dy}\right|_{y=y_0},
\ee
it is this {\it participating} matter which gives rise to the 
transverse flow signal.

This can also be seen by comparing the flow values derived 
from the transport calculation when using either all nucleons 
or cutting out the spectator nucleons, which can easily be identified 
by their collision number. Now defining the flow $F$ as
%$\langle \frac{dp_x}{dy} \rangle$ for 
derivative between 
$-0.5 \leq y/y_0 \leq 0.5$ we
obtain for a $Au+Au$ collisions at b=6 fm the results displayed in 
Table \ref{tab_flow}.
Clearly only at very low incident energies
($\sim$ 150 AMeV)
-- when the spectators move with relatively low velocities -- the results 
including or excluding spectator matter differ when calculating the flow.

\begin{table}[H]
\begin{center}
\begin{tabular}{c|c|c}
$E_{kin}$ & $F_{with}$ [MeV] & $F_{w/o}$ [MeV] \\
\hline\hline
150 AMeV &  66 &  86 \\
250 AMeV & 112 & 115 \\
400 AMeV & 138 & 139 \\
  1 AGeV & 197 & 197 \\
  2 AGeV & 222 & 221 
\end{tabular}
%\noindent {\bf Table \ref{tab_flow}:} 
\caption{
Flow values $\langle dp_x/dy \rangle$ for $Au+Au$ collisions at b=6 fm for
different incident energies as calculated within the CBUU model.
Center row: Including all particles, right row: Excluding particles 
with collision number zero.
}
\label{tab_flow}
\end{center}
\end{table}

Based on the observation, that sidewards flow is created by the
participating matter, it becomes transparent why flow
does not clearly distinguish between an EoS with and without
momentum-dependent forces. Since the fireball contains the
stopped matter, the relative momenta in the fireball
besides the unordered thermal motion,
are small. Only when applying additional
cuts, e.g. on high transverse momenta \cite{PD93,bass98}, 
i.e. by selecting particles escaping early from the fireball,
or selecting mainly participant or spectator particles 
by appropriate $\Theta_{cm}$-Cuts \cite{Cro97},
a difference between the momentum-dependent and 
momentum-independent EoS can be established.

\end{subsection}
%-------------------------------
\begin{subsection}{Conservation of angular momentum}

Since flow is generated by the participating matter
escaping from the fireball, where the particles move random 
and undergo numerous collisions,
the inclusion of an explicit angular 
momentum conservation mechanism for the two-body collisions 
in a transport model is of minor importance.
The conservation of angular momentum or at least the conservation of 
the reaction plane in the individual particle-particle collisions 
is usually neglected in transport models.
In \cite{GdG90,KK94} the effects of an explicit angular momentum
conservation mechanism have been studied, 
where in \cite{KK94} a substantial effect of a reaction plane 
conservation and a systematic choice of repulsive or
attractive scattering trajectories in the individual 
nucleon-nucleon collisions
on transverse flow is reported.
We, therefore, investigate the influence of modifications 
of the CBUU collision term on angular momentum conservation
and flow.
However, assuming both energy and angular 
momentum conservation in the individual particle-particle
collisions gives rise to some conceptional problems:
These two quantities determine the collision time
and angular distribution in a unique way.
The latter is usually simulated in line with the differential
cross section measurements for free $NN$ collisions and
thus should not be changed in favor of an arbitrary
collision geometry. The first, i.e. the time restriction, is
in contradiction to the concept of 
finite time steps and requires the relocation of the
colliding particles either in coordinate or momentum space
\cite{GdG90}.
This relocation again disturbs the evolution of the 
phase-space density and gives rise to an artificial motion 
of the particles.
To avoid these problems we, therefore, only
checked the importance of the conservation of the reaction plane, 
that is keeping the direction of $\vec L_{ij}$ constant
in the individual collisions between particles $i$ and $j$,
where similar to \cite{KK94} attractive or repulsive 
particle trajectories could be chosen.

As starting point in Fig.~\ref{L_t_Ni} the total angular momentum of
a semicentral (b=4 fm) $Ni+Ni$ system at different incident energies
is shown, where we have calculated 
\be
|L| = \frac{1}{N}\sum_{j=1}^{N} |l_j| \;,
\quad
|l_j| = \left(l_{jx}^2+l_{jy}^2+l_{jz}^2\right)^{1/2}
\vspace*{-1.5ex}
\ee
with
\be
\label{sum_i}
l_{jx}=\frac{1}{A}\sum_{i\in j} (r_z p_y - r_y p_z) \;,
\;l_{jy} = \dots \;,\; l_{jz}=\dots\;,
\ee
where $N$ denotes the number of testparticles per nucleon and 
$A$ is the mass number of the system. The sum over $i$ in 
Eq.~(\ref{sum_i}) runs over all particles (baryons and mesons) of the
ensemble $j$.
At 150 AMeV the angular momentum in the cascade mode is conserved better
than 2.6~\%; the mean field part again gives rise to some deviation 
from the initial value throughout the high density phase but in
total conserves angular momentum better than 0.2~\%.
The full calculation conserves $|L|$ by 4.7~\%.
At 2 AGeV the situation looks similar, but the deviations
are smaller. The cascade mode deviates by % from total conservation 
about 2.2~\%, the mean field mode about 0.2~\% and the 
full BUU mode about 3.2~\%.
These rather small deviations from total angular momentum conservation
already indicate a small influence of a different treatment of 
the collisions on the evolution of the system.

Now, including the conservation of the reaction plane in the individual 
collisions, 
the result is shown in Fig.~\ref{L_t_Ni_mod}, upper part.
The quality of total angular momentum conservation in the cascade mode
is the same, 2.6~\% compared to 2.7~\%; a similar behaviour is 
obtained for the full BUU mode.
Thus, for the total angular momentum the treatment of the individual
collisions seems to play a minor role.

In order to check a possibly more sensitive quantity, we have
additionally calculated the azimuthal distribution of 
particles in the forward hemisphere, both with 
and without a conservation of the reaction plane,
in the cascade mode for semicentral collisions of systems of different 
mass.
The results are shown in Fig.~\ref{dNdphi}. 
Whereas for a $pp$ collision the effects are expectedly dramatic,
already for the $dd$ system the signal is much weaker.
This is due to the fact that the collision plane of the
individual collisions is not identical to the reaction plane
of the system and more or less arbitrary. Consequently, with increasing 
system mass and collision number the effects of a reaction 
plane conservation in the individual collisions rapidly vanishes.
Table \ref{dNdphi_ratio} gives the ratio 
$R=N_0 / N_{90}$ obtained from a  $cos \phi+cos^2 \phi$ fit
to the data of Fig.~\ref{dNdphi}.

\begin{table}[H]
\begin{center}
\begin{tabular}{c|c|c|c|c}
 System & b [fm] & $R_{\rm with}$ & $R_{\rm w/o}$ & av. coll. number \\
\hline\hline
                pp & 0.2 & $\infty$       & $1.02\pm 0.02$ & 1 \\
                dd & 1   & $1.35\pm 0.02$ & $1.00\pm 0.01$ & 1,24 \\
 $^{4}$He $^{4}$He & 1   & $1.22\pm 0.02$ & $0.97\pm 0.02$ & 1,41 \\
 $^{12}$C $^{12}$C & 1.5 & $1.09\pm 0.02$ & $0.96\pm 0.02$ & 1,48 \\
 $^{16}$O $^{16}$O & 1.7 & $1.05\pm 0.01$ & $1.00\pm 0.01$ & 2,35 \\
$^{40}$Ca $^{40}$Ca& 2.3 & $1.03\pm 0.01$ & $0.98\pm 0.01$ & 2,95 
\end{tabular}
\end{center}
\caption{
Anisotropy ratio $R=N_0 / N_{90}$ at 150 AMeV for particles from the
forward hemisphere.
}
\label{dNdphi_ratio}
\end{table}

When increasing the incident energy above the one- and 
two-pion threshold, numerous inelastic collisions happen, 
where the angular momentum of the baryons cannot be conserved, anyway.
More important than the elastic scattering is the treatment of the
meson-nucleon-reactions here. Normally, when e.g. a pion 
matches the requirement
\be
b_{\pi N} \le \sqrt{\frac{\sigma_{\pi N}^{\rm tot}}{\pi}}
\ee
and also fulfils the Kodama-criterion \cite{kodama84},
it is absorbed by the nucleon and the created resonance is located
at the position of the absorbing nucleon.
Alternatively we apply a treatment, where the absorbing nucleon in a 
\mbox{$\pi N\to R$} (resonance) reaction is relocated to the 
center of momentum of the $\pi$ and the nucleon.
This procedure also does not conserve the angular momentum of 
the pion plus the nucleon relative to the lab system, but one can expect 
that the deviation 
%\be
%\Delta \vec l = \frac{\Delta \vec r_{\pi N} \times \Delta \vec p _{\pi N}}{2}
%\ee
is on average smaller than 
%\be
%\Delta \vec l = \Delta \vec r_{\pi N} \times \vec p _\pi 
%\ee
in the standard treatment.
In Fig.~\ref{L_t_Ni_mod}, lower part, 
the difference between the 'normal' BUU and this alternative treatment
is shown.
The total angular momentum conservation increases from 
2.2~\% to 1.2~\%. However, as in the discussion of the previous
paragraph, this relocation gives rise to unphysical sudden motion 
of the baryons which disturbs the time evolution of the system and which
we thus do normally not include in our model.

In summary, concerning sidewards flow, we find no systematic 
changes in the results of
$\langle \frac{dp_x}{dy} \rangle$, at least in the range
$-0.5 \le y/y_0 \le 0.5$,
within the statistical accuracy,
both for the $Ni+Ni$ and for the $Au+Au$ system at all energies
considered here (150 AMeV to 2 AGeV) depending on the treatment of the 
binary collisions.
The inclusion of a reaction plane conserving mechanism for the individual 
elastic particle-particle  collisions has some effect 
either on very small systems (cf. Fig.~\ref{dNdphi})
or when selecting particles with a low collision number in reactions
between heavier systems.

\end{subsection}
%-------------------------------
\begin{subsection}{Mass distribution of resonances}

Starting again from the observation that the participating matter 
is the origin of the transverse flow, another effect becomes
worth pointing out: The amount of flow at high energies 
depends on the mass distribution of the resonances, i.e.
on the number of nucleon resonances taken into account in a 
transport model and being excited in the high density phase.

Recent data on proton flow \cite{Herrmann96} indicate
a decrease of sidewards flow above 1 AGeV incident energy 
following the well known logarithmic increase at low energies.
Using standard potential parameterizations, both
nonrelativistic \cite{GBdG87,WPKdG88,wir88_2} and relativistic 
\cite{maru94,pradip}, this behavior cannot be understood within 
conventional transport models.
In the latter the optical potential stays constant or even increases 
at high momenta and therefore the repulsion generated
from the momentum-dependent forces in a HIC gives rise to a 
significant contribution to the flow signal.
However, since the nucleon-nucleus optical potential is 
only known up to 1 GeV experimentally \cite{Hama}, 
it was recently proposed that this decrease of flow
above 1 AGeV might indicate a decrease of the optical 
potential at high relative momenta or at high baryon density 
\cite{pradip}.

In \cite{BKL91,BKW89} the transverse momentum of the baryons 
has been disentangled into a collisional part, a mean-field part 
and a part originating from the Fermi-motion of the particles,
i.e.
\be
p_t = p_t^{\rm coll} + p_t^{\rm MF} + p_t^{\rm Fermi} \;.
\ee
Depending on the EoS and rapidity interval considered, 
the relative contributions of these vary between
27, 73 and 0 \% for $y\simeq 0.5\,y_{pro}$ and 
37, 0 and 63 \% for $y > y_{pro}$ for intermediate impact parameters.
For more central collisions the Fermi-part contributes 
at minimum about 25 \% over all rapidities.
However, since we deal with intermediate impact parameters
$b\simeq 0.5\,b_{max}$ only, which correspond to the 
multiplicity bins M3 and M4 where maximum transverse flow
is produced, we neglect in the following the contribution
of the Fermi-motion to the flow.

Thus disentangling the flow signal into a collisional and
a potential part, it turns out that $\sim$50 \% of the
flow stems from the particle-particle collisions
while roughly another 50 \% are generated by the potential repulsion.
Figure~\ref{trfl_eos} shows these two contributions for flow
for the system $Ni+Ni$ at b=4 fm in comparison to the 
experimental data as compiled by Ref.~\cite{Herrmann96} 
using different EoS denoted by 'hard', 'medium' and
'soft'. Both the collisional ''background'' 
and the potential part rise up to 1 AGeV incident energy 
and remain constant above, whereas the data indicate a 
decrease above 1 AGeV. As seen from
Figure~\ref{trfl_eos}, the $Ni+Ni$ data are described reasonably by both 
a 'soft' and 'medium' EoS.

Fig.~\ref{trfl_mres}, furthermore, shows in the upper part
the flow in the cascade mode of the CBUU model, 
i.e. the collisional part, for the $Ni+Ni$ system 
with all resonances up to M=1.95 GeV included (solid line) 
and without the resonances above the $\Delta(1232)$ (dashed line).
In the latter case all inelastic scattering strength
is put into the $NN\to N\Delta$-channel, thus no stopping
power from the inelastic collisions is lost. It is clearly seen that the
excitation of higher nucleon resonances, though quite low in number, 
leads to an enhanced flow above 1 AGeV.
In the lower part of Fig. \ref{trfl_mres} the results of 
the RBUU model \cite{kweber93,maru94,KLW87,BKM93,kweber92}, 
where only the $\Delta(1232)$ is included, is shown by the solid line.
The flow results practically coincide with those from the CBUU model
when including only the $\Delta$ resonance. 
In addition, we have calculated the flow within the RBUU model
by shifting the $\Delta$ mass artificially to 1500 MeV (dashed line) in
order to simulate effects from a heavy resonance. Again the flow
increases compared to a calculation with the $\Delta(1232)$ mass above
1 AGeV and thus demonstrates the sensitivity of this observable with
respect to the inelastic channels taken into account.

We have analysed the origin of flow as a function of time within the
various models and found that this effect is caused mainly by two reasons:
(a) the Pauli blocking of final scattering states in the 
initial phase of the reaction, which
is reduced in case of high mass excitations,
and (b) a harder meson spectrum
emitted by the fireball in case of higher resonances since these
mesons partly are absorbed by the spectator matter which achieves a larger
momentum transfer. Thus high mass resonances have to be included in any
transport study that attempts to draw conclusions on the EoS or momentum
dependence of the optical potential in comparison to experimental flow data
above bombarding energies of about 1 AGeV.

\end{subsection}
\end{section}

%% file: radfl.tex
\begin{section}{Radial flow}
\label{rad_flow}

Radial flow has been discovered \cite{Jeong94}
when analyzing the flow pattern of very central events of HIC.
In contrast to transverse flow up to about 70 \% of the
incident energy (stored in the hot compressed fireball)
is released as ordered radial expansion of the 
nuclear matter. Thus the hope is to extract information
especially on the compressibility of the EoS via the magnitude of the 
radial flow.

%-------------------------------
\begin{subsection}{Temperature and flow energy}

Experimentally the radial flow is characterized or fitted in terms of
the Siemens-Rasmussen formula 
\cite{SR79}
\be
\label{SR}
\frac{d^3N}{dEd^2\Omega} \sim
p \cdot e^{-\gamma E/T} 
\left\{\frac{\sinh \alpha}{\alpha} \cdot (\gamma E + T) 
- T \cdot \cosh\alpha \right\}
\ee
with $\gamma=(1-\beta^2)^{-1/2}$ and $\alpha=\gamma\beta p/T$,
where $\beta$ denotes the flow velocity and $T$ characterizes
some temperature. We follow the same strategy in our CBUU calculations
and apply a least square fit to the CBUU nucleon spectra using 
Eq.~(\ref{SR}).

The results of the CBUU calculations for 
central $Au+Au$ collisions are shown in Fig.~\ref{rad} as a function 
of the bombarding energy 
in comparison to the data from \cite{Lisa95, Poggi95, Hong97}.
We find that the 'temperature' $T$ is systematically underpredicted 
in all schemes investigated 
%('s': soft EoS; 'h'; hard EoS; 'smd', 'mmd' and
%'hmd': soft, medium and hard momentum dependent EoS, respectively),
(soft and hard EoS, with and without momentum dependent forces),
and that the flow velocities are not correctly reproduced, being too
low at low energy and crossing the experimental data around 800 AMeV.
This might, on the one side, indicate a strong binding from 
the potential, which gives not enough repulsion at high densities and
overcompensates the collisional pressure from the fireball.
However, on the other side, the nucleon spectra resulting 
from the CBUU calculation show a strong non-thermal component 
at low incident energies 
and are thus in contradiction to the physical picture behind
Eq.~(\ref{SR}) which assumes an isentropic expansion of a
thermal equilibrated source.
In a recent publication \cite{hom98_1} we have investigated the degree 
of equilibration in a HIC as a function of the incident energy and
the system mass and have found that even the most massive systems
like $Au+Au$ do not equilibrate at low energies. 
This is also in line with earlier findings \cite{LBK90}.

Thus we also evaluate the radial flow energy according to a prescription
used in \cite{LiKo97,NP97},
\be 
\label{NP1}
E_{flow} = \frac{1}{A}\sum_{i=1}^{A}
\left(\sqrt{m_i^2+\frac{(\vec r_i \cdot \vec p_i)^2}{r_i^2}}
-m_i\right) \;,
\ee
which then gives the thermal energy as the difference to the 
total kinetic energy of the system
\be
\label{NP2}
E_{therm} = \frac{1}{A}\sum_{i=1}^{A}
\left(\sqrt{m_i^2+p_i^2}-m_i\right) - E_{flow} \;.
\ee
This direct evaluation of $E_{flow}$ and $E_{therm}$
or $\beta$ and $T$, respectively, has the additional advantage
of being statistically more stable than the fits to the nucleon 
spectra using Eq.~(\ref{SR}).

However, although Eq.~(\ref{NP1}) obviously yields the radial motion 
of the system, the resulting flow values $\beta_{radial}$  
differ strongly from the ones using Eq.~(\ref{SR}) as shown in 
Fig.~\ref{rad_b_SRNP} for central $Au+Au$ collisions. 
Additionally, the results depend strongly on the evaluation time, 
since particle momentum and coordinate 
position become more and more aligned throughout the 
expansion phase. This is shown in Fig.~\ref{rad_t_SRNP} for a 
central $Au+Au$ collision at 1 AGeV for the parameter $T$ 
as well as the parameter $\beta_{flow}$.
The changes of the results with time when using 
Eq.~(\ref{SR}) are much less pronounced.

Consequently, one might ask for an ''optimal'' time for
evaluating these quantities, at least when using 
Eqs.~(\ref{NP1},\ref{NP2}).
Looking at the time evolution of the central density, 
collision number and $\pi,\Delta$ - abundancy throughout a central 
$Au+Au$ collision at 1 AGeV (Fig.~\ref{gyuri}), we find that for 
this reaction the collision rate drops practically to zero at
t $\simeq$ 25 fm/c.
This is also the time when the average expansion velocities
in transverse and longitudinal direction become approximately the
same as shown in Fig.~\ref{allbeta} and the velocity profile
reaches its final shape (Fig.~\ref{beta_x_t}).
Thus, inspite of further pion production via resonance decays,
the reaction dynamics determining the flow is basically over at 
t $\approx$ 25 fm/c and this time might be considered as a kind of
hadronic freeze-out time. 
Nevertheless, the motion of the baryons becomes more 
ordered during the further expansion process as
can be seen in Fig.~\ref{allbeta}. Consequently,
when using Eq.~(\ref{NP1}) $E_{flow}$ rises with
time while $E_{therm}$ drops with time.
However, already at the 'freeze-out time' t $\leq$ 25 fm/c the
results using Eqs.~(\ref{NP1},\ref{NP2}) and Eq.~(\ref{SR})
differ quite substantially.

Thus, in addition to Eq.~(\ref{SR}) and Eq.~(\ref{NP2})
for evaluating the 'temperature' $T$ we have determined a 'temperature'
via the $N/\Delta$ - ratio. 
Since in \cite{hom98_1} we have shown that at 1 AGeV at least 
isospin-equilibration is reached in a $Au+Au$ collision,
we assume that also local thermal equilibrium is achieved 
and the $N/\Delta$ - ratio might give a hint to the 'real' 
temperature of the system.
The temperature $T$ can be determined from
\be
\frac{N_\Delta}{N_N} \,=\, 
\frac{16}{4} \, 
\frac{\int dM \, f(M) \, M^{3/2} \, e^{-E(p,M)/T} }
{ m^{3/2} \, e^{-E(p)/T} }\;,
\ee
where $m,M$ denote the nucleon and $\Delta$ mass, respectively.
The factor 16/4 comes from spin-isosopin and
\be
f(M) \,=\, \frac{2}{\pi}\,
\frac{M^2 \Gamma(M)}{(M^2-M_R^2)^2 + M^2\Gamma^2(M)}
\ee
is the normalized $\Delta$-mass distribution using
\be
\Gamma(M) \equiv \Gamma(p) \,=\, 
\Gamma_R \frac{M_R}{M} \left(\frac{p}{p_R}\right)^3 
\left(\frac{p_R^2 + \delta^2}{p^2 + \delta^2}\right)^2 \quad,
\ee
as the momentum dependent $\Delta$-width with
$\delta^2 = (M_R - m_N - m_\pi)^2 + \Gamma_R^2/4$, 
$\Gamma_R$=150 MeV and $p$ denoting the pion momentum in
the Delta rest frame.
Fig.~\ref{T_t_SRNPND} shows the corresponding temperature
in terms of the dotted line in comparison to the temperatures obtained using 
Eqs.~(\ref{SR}) (full line) and (\ref{NP2}) (dashed line) as a function
of time again for $Au+Au$ at 1 AGeV.
Though the $N/\Delta$-temperature shows a similar 
time dependence as the temperature obtained from Eq.~(\ref{NP2}),
its actual values are closer to the temperatures obtained 
from the nucleon spectra over the whole time range.
Especially at t=25 fm/c, where the collision rate 
drops to zero and the central density in the
reaction volume gets below $\rho_0$, it is practically
identical to the temperature from Eq.~(\ref{SR}).
Thus one might define t $\approx$ 25 fm/c as the freeze-out
time for this particular reaction. 
The nucleon spectra parameter $T$ then conserves the $N/\Delta$ -
'temperature' at this time.

\end{subsection}
%-------------------------------
\end{section}

%% file: mass.tex
\begin{section}{Mass dependence}
\label{mass}

As reported in Refs.~\cite{Ritter94,Partlan95,RR97}
flow signals, both radial and transversal,
are more pronounced for fragments with charge Z=2,3,$\cdots$ 
than for free protons or neutrons. 
This is usually attributed to i) the superposition of the
ordered collective flow with the unordered thermal motion
of the particles and ii) to the subsequent decay of 
excited heavier fragments during the expansion phase of 
the HIC into fragment plus nucleons: $F_{(A+n)}' \rightarrow F_{(A)} + n\times N$.
Thus a direct comparison of the results from a 
single-particle model like CBUU, which describes the
time evolution of the {\it single-particle} 
phase-space  density, to the experimental data averaged over 
different fragments \cite{Lisa95} seems questionable. 
We thus  study how far one
can come in describing fragment flow data within the single-particle
approach and then estimate the role of many-body correlations by the
remaining deficiences.
For transverse flow this approach has been used earlier 
\cite{KBC90,KBC91} in comparison to the experimental data 
from Refs.~\cite{kam89,GPR89}.

We have applied a simple phase-space coalescence model
to the final state distribution of the testparticles 
to form fragments out of the single-particle distribution. In this respect
particles are decided to belong to a fragment, if their relative
distance is less than $r_0$ in coordinate space and smaller than 
$p_0$ in momentum space. While $p_0$ was fixed to 250 MeV, 
which roughly corresponds
to the average Fermi momentum of light nuclei, 
$r_0$ was chosen to reproduce the experimentally measured 
number of free nucleons in the collision.

The results for a central 1 AGeV Au+Au collision is shown in 
Fig.~\ref{massdep_radflow} for both the temperature $T$
and flow velocity $\beta_{radial}$ derived from the respective 
spectra using Eq.~(\ref{SR})
in comparison to the results of Ref. \cite{Lisa95} for
proton spectra and averaged fits. In addition, the
CBUU results from fitting the pure single-particle spectra 
are also shown in terms of the full squares.
As can be seen from the upper picture, when applying the 
coalescence model to the final single-particle distribution,
the temperature $T$ for the remaining
nucleons starts at the single-particle temperature (full square)
and then increases with 
fragment mass towards the averaged ex\-peri\-mental 
value. This increase is stronger for a hard equation of state
(denoted by 'h' or 'hmd', respectively) as for a soft EoS ('s' and 'smd',
respectively) due to a stronger collectivity of the nuclear motion.
Averaging over the CBUU results for the different masses,
the resulting mean temperature is in good agreement with 
the averaged ex\-peri\-men\-tal measurement while the 
single-particle temperature is close to the ex\-peri\-men\-tal
value for protons. 

The situation is similar for the flow velocity which drops with 
increasing mass, again roughly in line with the experimental 
findings of Ref. \cite{Lisa95}.
Also here the average over the CBUU results in the mass range 
of A = 1--4  leads to a flow value closer to the averaged experimental 
value. 
The different EoS employed (hard/soft, with and without momentum-dependent
forces) practically lead to the same result for the flow velocity
$\beta_{radial}$.

Again the results are quantitatively different when using Eq.~(\ref{NP1})
for evaluating the flow velocity. Whereas from the 
analysis of the spectra no difference between EoS of different
compressibilities can be inferred, the flow energy 
derived using Eq.~(\ref{NP1}) clearly separates a hard
and soft EoS. Figure~\ref{cl2_Eflow} shows the results of 
the CBUU calculation for the 1 AGeV Au+Au collision 
for fragment masses of 1 to 4 in comparison to the data from 
\cite{Lisa95}. This comparison favors a hard EoS (full squares).

For transverse flow we have compared the flow angle \cite{RR97},
\be
\label{angle}
\Theta_S = \tan^{-1}\left(\frac{d\langle u_x \rangle}{d u_z}\right)\;,
\ee
with $u_x$ and $u_z$ denoting the velocity in transverse and
longitudinal direction, respectively,
to the data from Ref.~\cite{RR97}. Since the
statistics of the transverse momentum distributions
of the fragments obtained by the coalescence model
is poor, we have evaluated Eq.~(\ref{angle}) for a given mass
without separating for charge additionally. To be comparable 
to Ref.~\cite{RR97}, we estimate the mass number of the
different charges measured by multiplying the given 
charge number 
with the average mass number for that charge.
The result is shown in Figure~\ref{theta_s} for a Au+Au
collision at 400 AMeV in terms of the full squares whereas the data
are given in terms of the open circles.
For the higher masses \mbox{(A $\ge$ 4)} the agreement between
the data and the CBUU calculation is very good, however,
in the low mass region the flow angle is overestimated 
in the CBUU plus coalescence model quite significantly.

In summary, the application of a simple coalescence model 
can provide results about the mass dependence 
of the radial flow parameters $T$ and $\beta_{radial}$ that are
both correct in the functional dependence on mass 
and in their absolute magnitude. Here the important 
aspect is the suppression of  
random thermal motion, which can be provided by
an arbitrary clustering algorithm.
However, for transverse flow the situation is different.
Obviously the experimental proton momentum distribution
is heavily distorted by massive fragment (or spectator)
decay that is not adequately described in the single-particle model.
This contrasts to the findings in an earlier
publication \cite{KBC90} which compared to PlasticBall
data \cite{kam89,GPR89}.

\end{section}

%% file: summry.tex
\begin{section}{Summary}
\label{summary}

In this paper we have explored the dependence of transverse and radial
flow signals on various model inputs and evaluation prescriptions 
using the CBUU model.

For the transverse flow its origin could be traced back to
the expanding participating matter. Starting from this observation
we investigated the influence of different EoS, the conservation
of angular momentum in the individual particle-particle collisions 
and of the mass distribution of the nucleonic resonances.

The most important point is that in general angular momentum 
conservation in the individual particle-particle collisions 
can be neglected for the overall description of a HIC. 
In contrast to that the mass distribution of the resonances
included in the model -- though the high mass resonances are 
quite low in number throughout a HIC -- 
plays an important role for the description of transverse flow
above \mbox{1 AGeV}.
Since they change the collisional part of the signal substantially,
they also influence the conclusions that one might draw from the
experimental flow measurements on the nuclear potential and the EoS.

For the radial flow we have concentrated on the difference between the
results for the flow temperature $T$ and flow velocity $\beta$
when using different EoS and different evaluation prescriptions.
The latter we see as an additional problem to the dependence 
of calculational results on nonphysical model inputs 
as recently pointed out also by Hartnack et al. \cite{Hartnack98}.
We showed that some kind of 'hadronic freeze out time' can be 
defined when looking at the collision rate, the isotropy of expansion 
and the flow velocity profiles and that at least at energies 
around 1 AGeV and heavy systems the thermodynamic temperature 
given by the $N/\Delta$-ratio at this time is best reflected by the
'temperature' of the nucleon spectra.

In a third part we have investigated the fragment-mass dependence
of the flow signals and have shown that even a simple coalescence 
model can provide reasonable agreement with the experimental
findings. From this one might conclude that at least 
in central reactions of heavy systems \mbox{($\to$ radial flow)}
we have a thermal scenario around \mbox{1 AGeV}, 
while the difference between model and experiment for 
semicentral events ($\to$ transverse flow)
hints towards strong nonthermal contributions to the signals measured.

\end{section}

%% file: captions.tex
\pagebreak
\pagestyle{empty}
\begin{section}*{Figure captions}

\noindent
{\bf Fig.~\ref{ener_t_Au}}:
Total energy per nucleon (nucleon restmass subtracted) for a central
$Au+Au$ collision at 1 AGeV. Upper part: cascade mode;
middle part: Mean-field propagation; lower part:
full calculation including mean-field propagation and collisions.\\

\noindent
{\bf Fig.~\ref{au1b6movie}}:
Coordinate space picture of a $Au+Au$ collision at 1 AGeV 
at b=6 fm for different times. Initial stage (upper left), 
intermediate (upper right) and final stages (lower plots). 
The contour lines represent densities of
0.1, 0.5, 1, 1.5 and 2 times $\rho_0$; the arrows indicate 
the direction and velocity of motion. In the lower plots 
the participating matter escaping from the reaction zone
circumfloats the spectators.\\

\noindent
{\bf Fig.~\ref{L_t_Ni}}:
Angular momentum per nucleon in a b=4 fm $Ni+Ni$ collision
at 150 AMeV (upper part) and 2 AGeV (lower part). 
Shown are the cascade mode (dotted line), Vlasov mode
(dashed line) and the full-BUU results (solid line).\\

\noindent
{\bf Fig.~\ref{L_t_Ni_mod}}:
Angular momentum per nucleon in a b=4 fm $Ni+Ni$ collision
in the cascade mode.
Upper part: Normal mode (dotted line) and with
reaction plane conservation in the individual baryon-baryon 
collisions (solid line) at 150 AMeV.
Lower part: Normal mode (dotted line) and with 
relocation of the absorbing baryon to the CM in a meson-nucleon
reaction (solid line) at 2 AGeV.\\

\noindent
{\bf Fig.~\ref{dNdphi}}:
Azimuthal distribution of particles emitted into the forward
hemisphere for systems of different size colliding with
${\rm b}\sim 0.6\times {\rm b}_{max}$ where ${\rm b}_{max}$ 
is $2 \times 1.125 A^{1/3}$. The solid and dashed histograms 
are the CBUU results, the lines in the lower 3 panels
are the appropriate fits in $\cos^2(\phi)$ leading to the
ratios of Table~\ref{dNdphi_ratio}.\\

\noindent
{\bf Fig.~\ref{trfl_eos}}:
Transverse flow for a $Ni+Ni$ collision at b=4 fm 
as calculated within the CBUU model in cascade mode
(dashed lower line) and for equations of state with 
different compressibilities  $K$
in comparison to data from EOS, Plastic Ball and FOPI
as compiled by Ref.~\cite{Herrmann96}.\\

\noindent
{\bf Fig.~\ref{trfl_mres}}:
Transverse flow in the cascade mode from
different transport models, CBUU (upper part) and RBUU 
\cite{kweber92,kweber93,maru94} (lower part).
The computations have been performed with different numbers of 
baryon resonances and resonance properties (see text). \\

\noindent
{\bf Fig.~\ref{rad}}:
The radial flow velocity (lower part)
and temperature (upper part) for central $Au+Au$ collisions 
evaluated via Eq.~(\ref{SR}) from the CBUU calculations in comparison
to the experimental data from Refs.~\cite{Lisa95,Poggi95,Hong97}.
The symbol 's' denotes a soft EoS without momentum dependent forces,
'h' a hard EoS and 'smd', 'mmd' and 'hmd' correspond to a soft, 
medium and hard momentum dependent EoS, respectively.\\

\noindent
{\bf Fig.~\ref{rad_b_SRNP}}:
The radial flow velocity 
for central $Au+Au$ collisions evaluated according
to Eq.~(\ref{SR}) and Eq.~(\ref{NP2}) from the CBUU calculations
as a function of the bombarding energy per nucleon $E/A$.\\

\noindent
{\bf Fig.~\ref{rad_t_SRNP}}:
The radial flow velocity (upper part) and temperature 
(lower part) for a $Au+Au$ collision at 1 AGeV evaluated according
to Eq.~(\ref{SR}) and Eqs.~(\ref{NP1},\ref{NP2})
as a function of the reaction time.\\

\noindent
{\bf Fig.~\ref{gyuri}}:
Time evolution of the central density, collision rate and 
pion number throughout a central $Au+Au$ collision at 1 AGeV.
The thin vertical line marks t=25 fm/c.\\

\noindent
{\bf Fig.~\ref{allbeta}}:
Time evolution of the total and radial mean velocity
in transverse and longitudinal direction
throughout a central $Au+Au$ collision at 1 AGeV.
The total velocity is given by
$\frac{1}{A} \sum_i \frac{|\vec p_i|}{E_i}$, the radial velocity by
$\frac{1}{A} \sum_i \frac{(\vec p_i\vec r_i)}{E_i|\vec r_i|}$.\\

\noindent
{\bf Fig.~\ref{beta_x_t}}:
The radial transverse velocity profile
at different times of a
central $Au+Au$ collision at 1 AGeV. The final shape of the
distribution is reached between t=22.5 and 25 fm/c.\\

\noindent
{\bf Fig.~\ref{T_t_SRNPND}}:
Temperatures obtained from the analysis of nucleon spectra
using Eq.~(\ref{SR}) (solid squares); direct evaluation
according to Eqs.~(\ref{NP1},\ref{NP2}) (open squares)
and from the $N/\Delta$ ratio (dotted line).\\

\noindent
{\bf Fig.~\ref{massdep_radflow}}:
Fragment mass dependence of the parameters temperature $T$ and 
flow velocity $\beta$ from particle spectra using 
the CBUU plus coalescence model. 
Also shown are the results for the fit to the 
CBUU single-particle spectra (s.p., left) and the
fits to the experimental spectra from Ref.~\cite{Lisa95}
(exp., right).\\

\noindent
{\bf Fig.~\ref{cl2_Eflow}}:
Flow energy for fragments of different mass evaluated
using Eq.~(\ref{NP1}) for a central 1 AGeV $Au+Au$ collision.
The experimental data are taken from Ref.~\cite{Lisa95}.\\

\noindent
{\bf Fig.~\ref{theta_s}}:
Transverse flow angles for different mass fragments in a 
400 AMeV $Au+Au$ collision using the CBUU plus coalescence model
in comparison to the data from Ref.~\cite{RR97}.
The horizontal lines indicate the flow angle saturation value.\\

\end{section}

%% file: figures.tex
\pagestyle{empty}
\begin{section}*{}

\begin{figure}[H]
\vspace*{-1cm}
\begin{center}
\includegraphics[angle=90,totalheight=22cm]{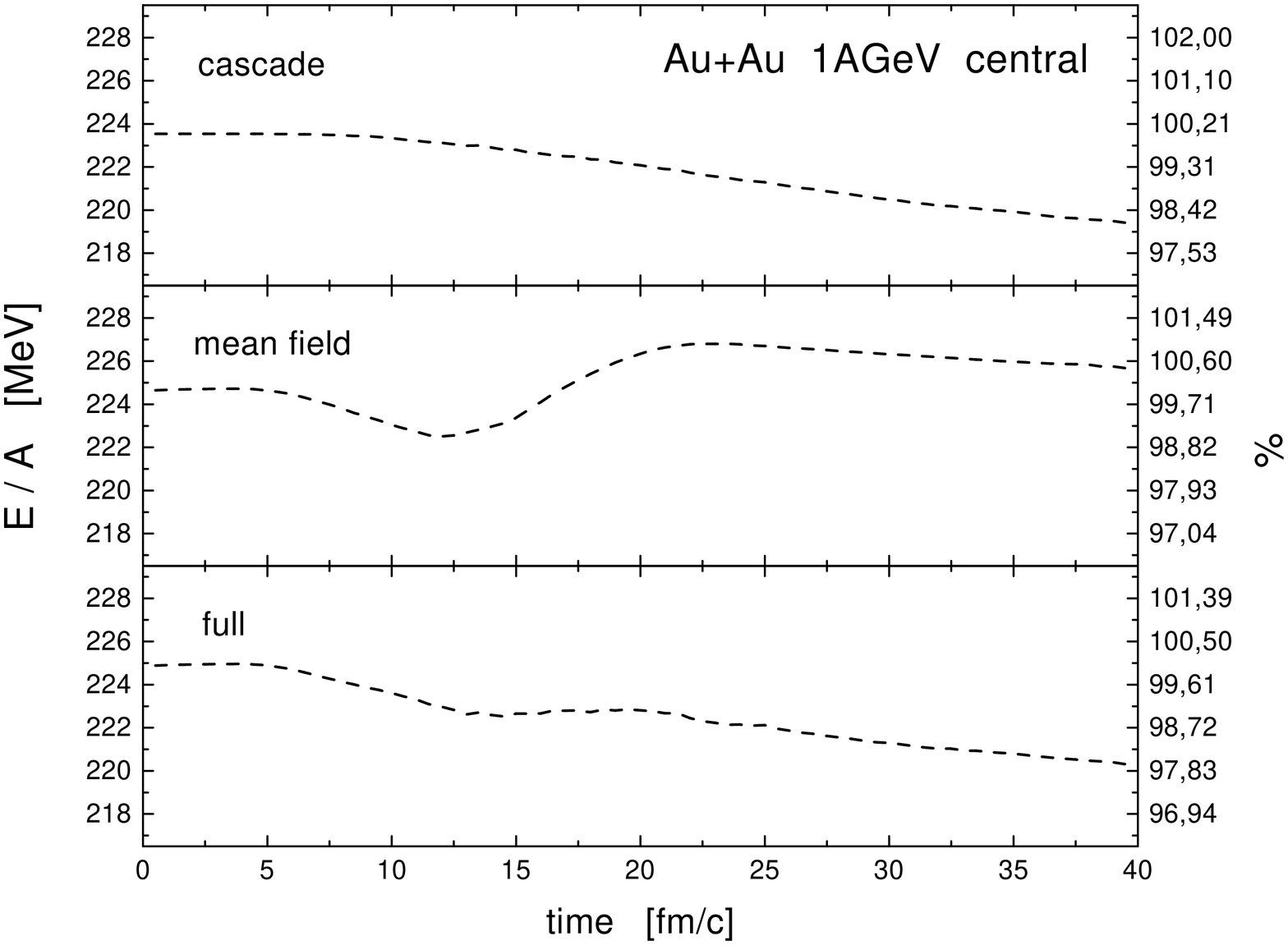}
\end{center}
\vspace{1cm}
\caption{}
\label{ener_t_Au}
\end{figure}

\begin{figure}[H]
\begin{center}
\hspace{-2.5cm}
\includegraphics[totalheight=7.5cm]{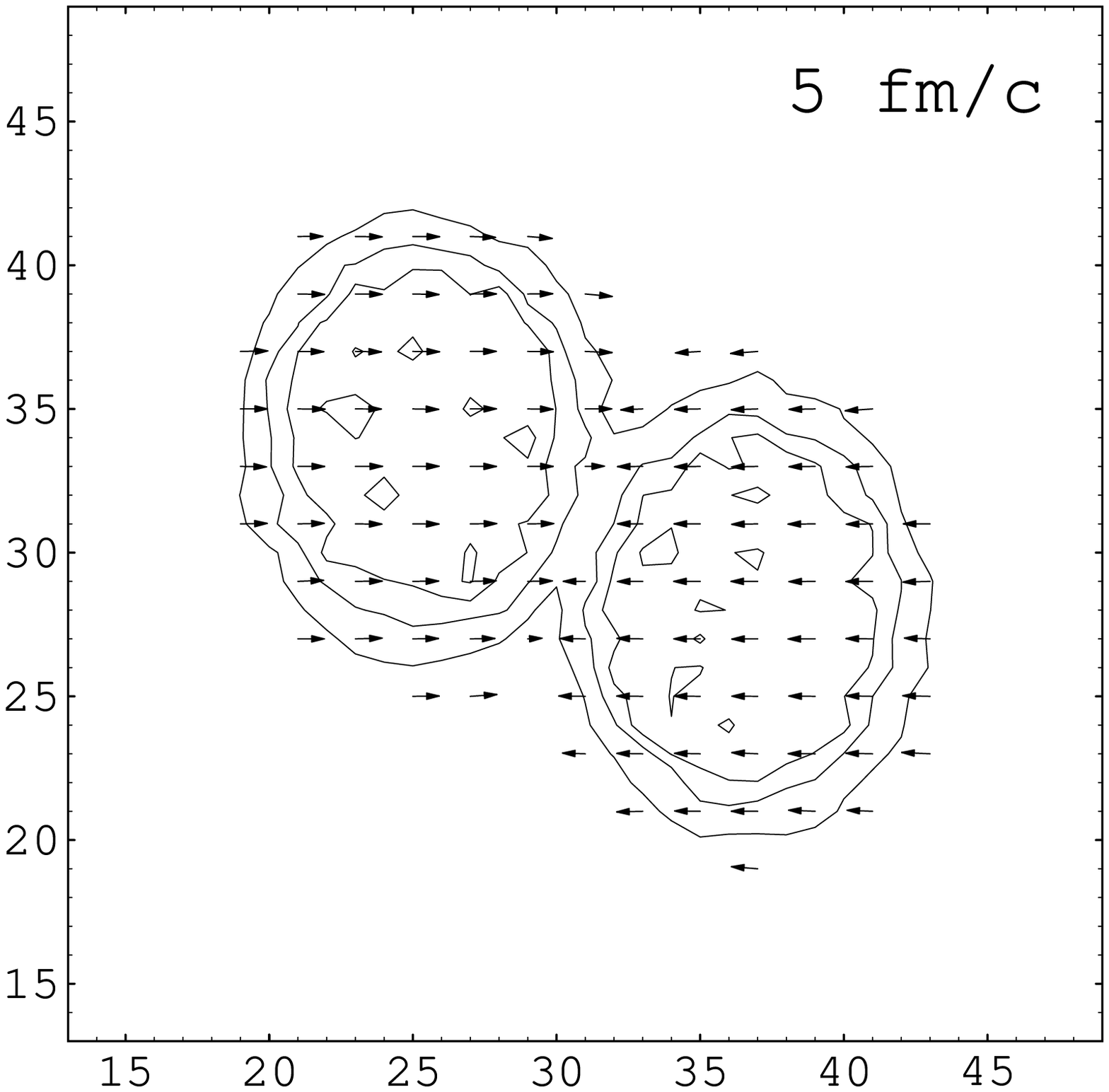}
\hspace{0.25cm}
\includegraphics[totalheight=7.5cm]{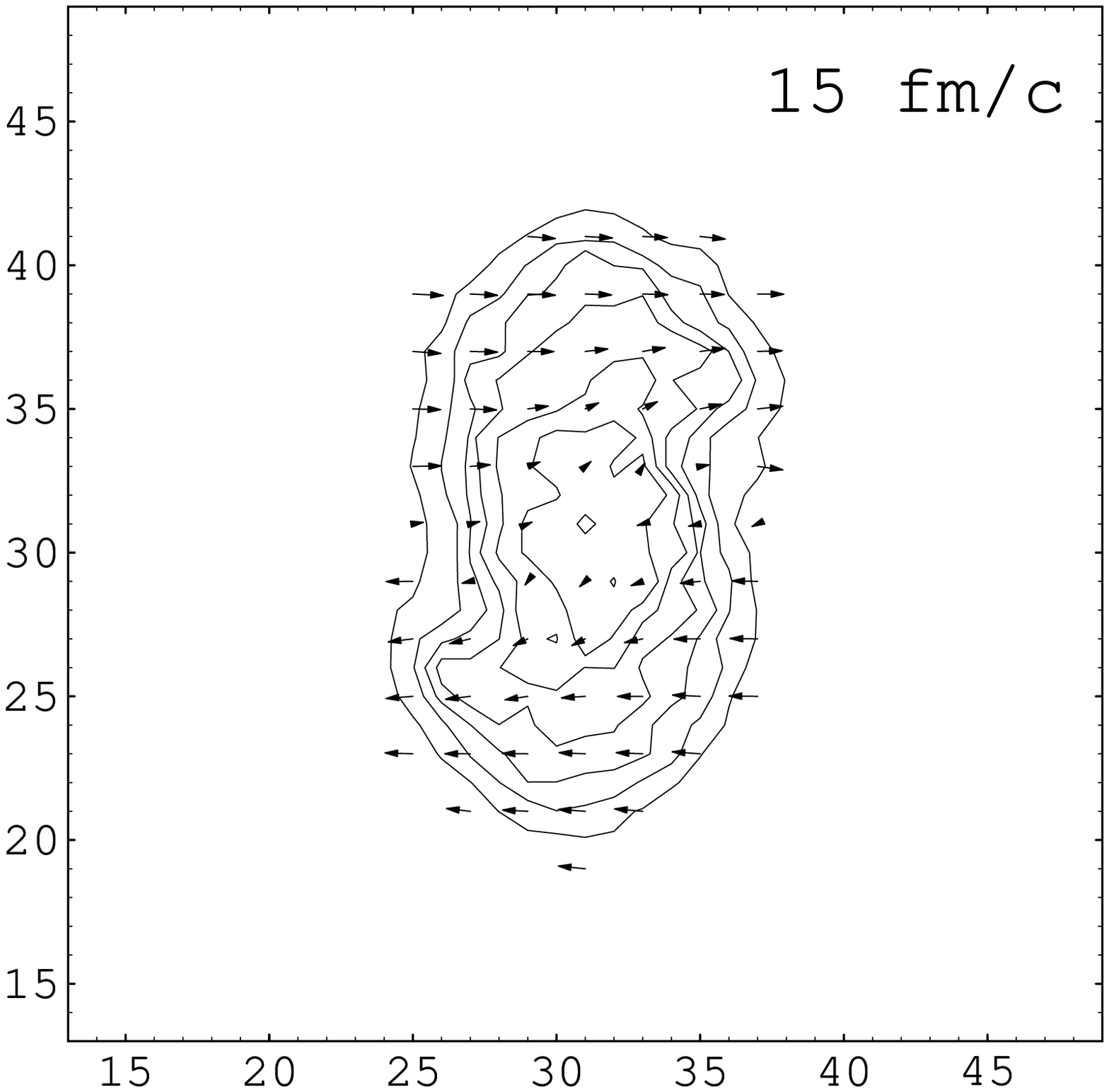}\\
\vspace{0.25cm}
\hspace{-2.5cm}
\includegraphics[totalheight=7.5cm]{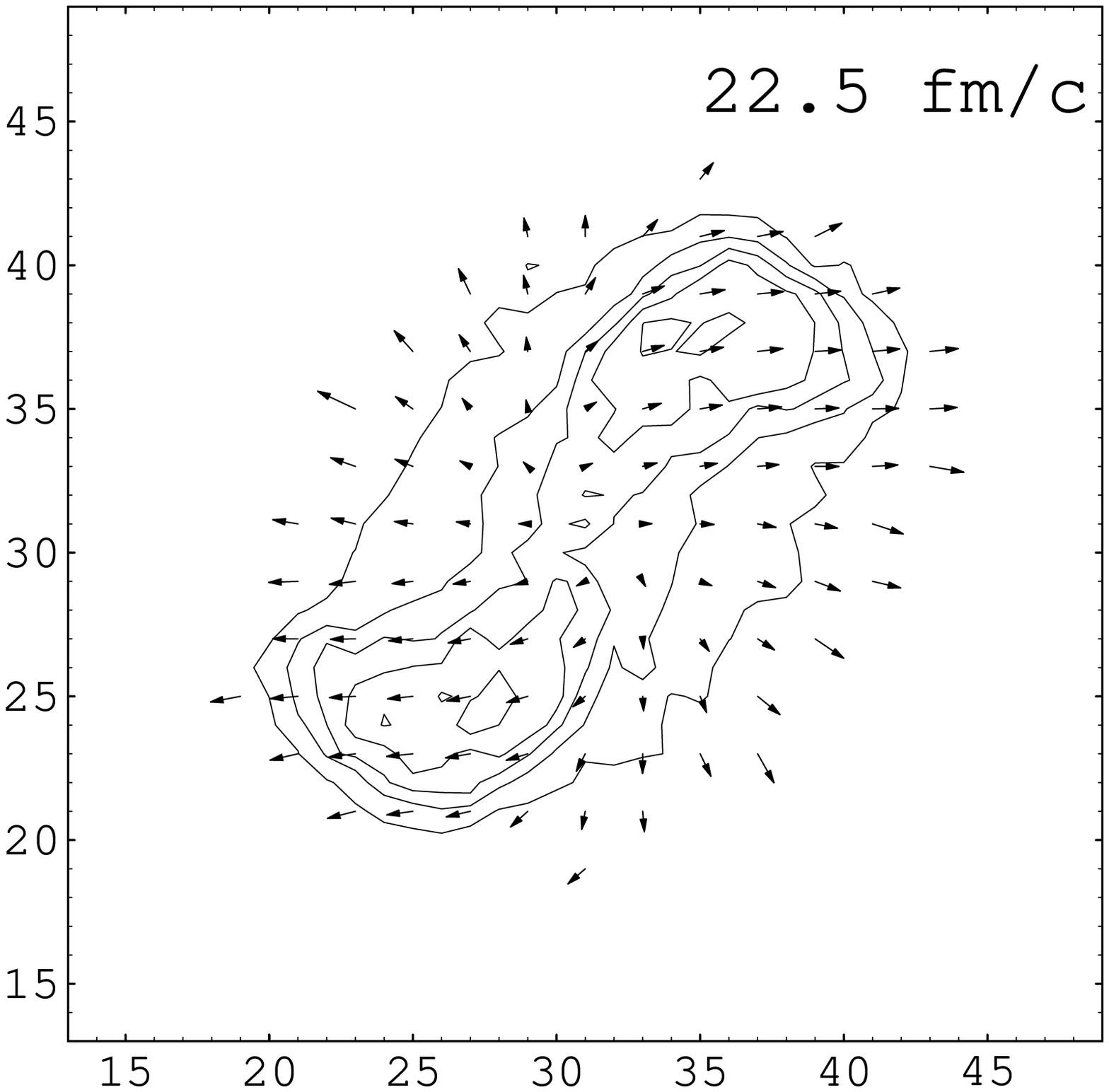}
\hspace{0.25cm}
\includegraphics[totalheight=7.5cm]{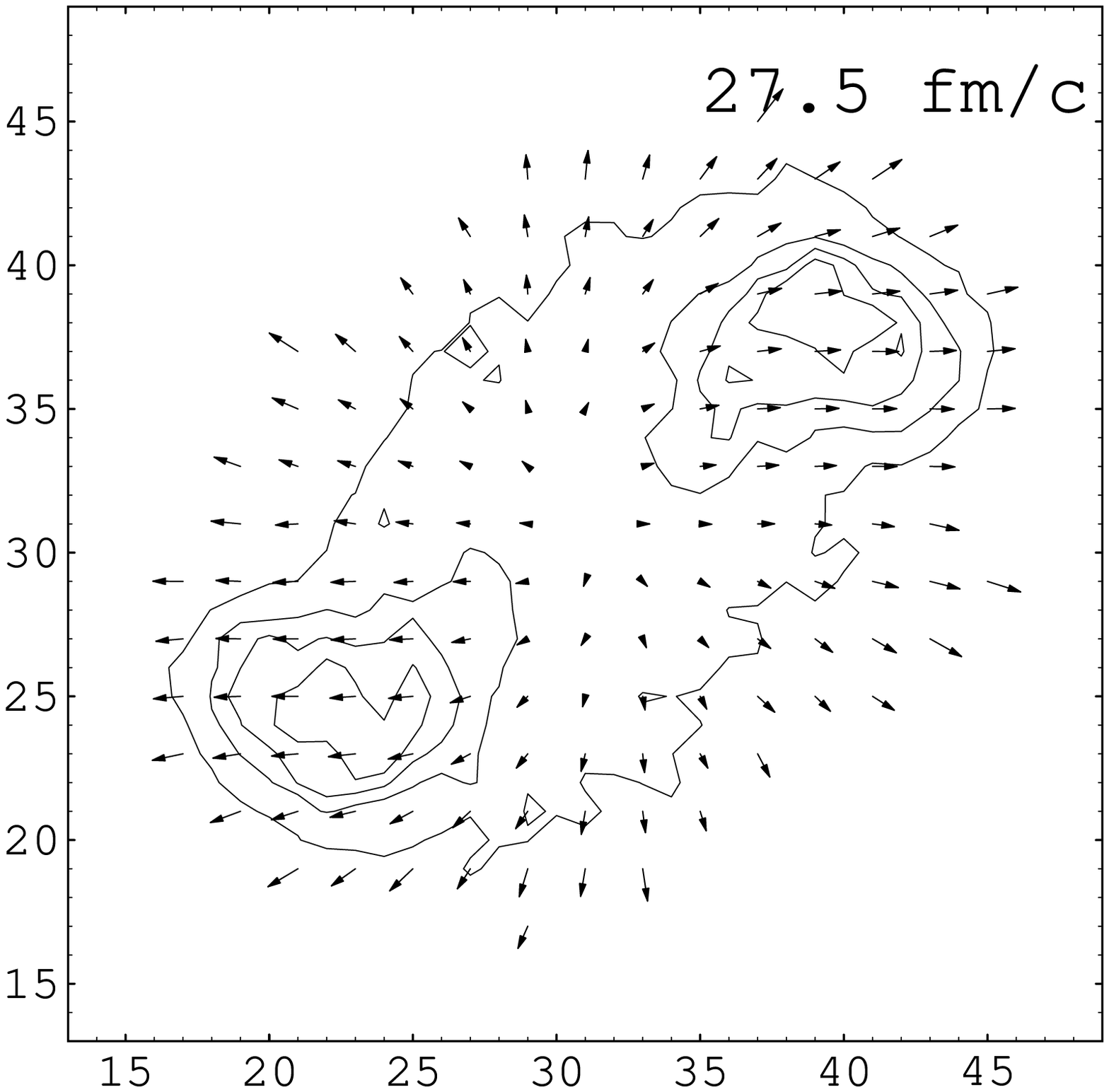}
\end{center}
\vspace{5cm}
\caption{}
\label{au1b6movie}
\end{figure}

\begin{figure}[H]
\begin{center}
\includegraphics[angle=90,totalheight=21cm]{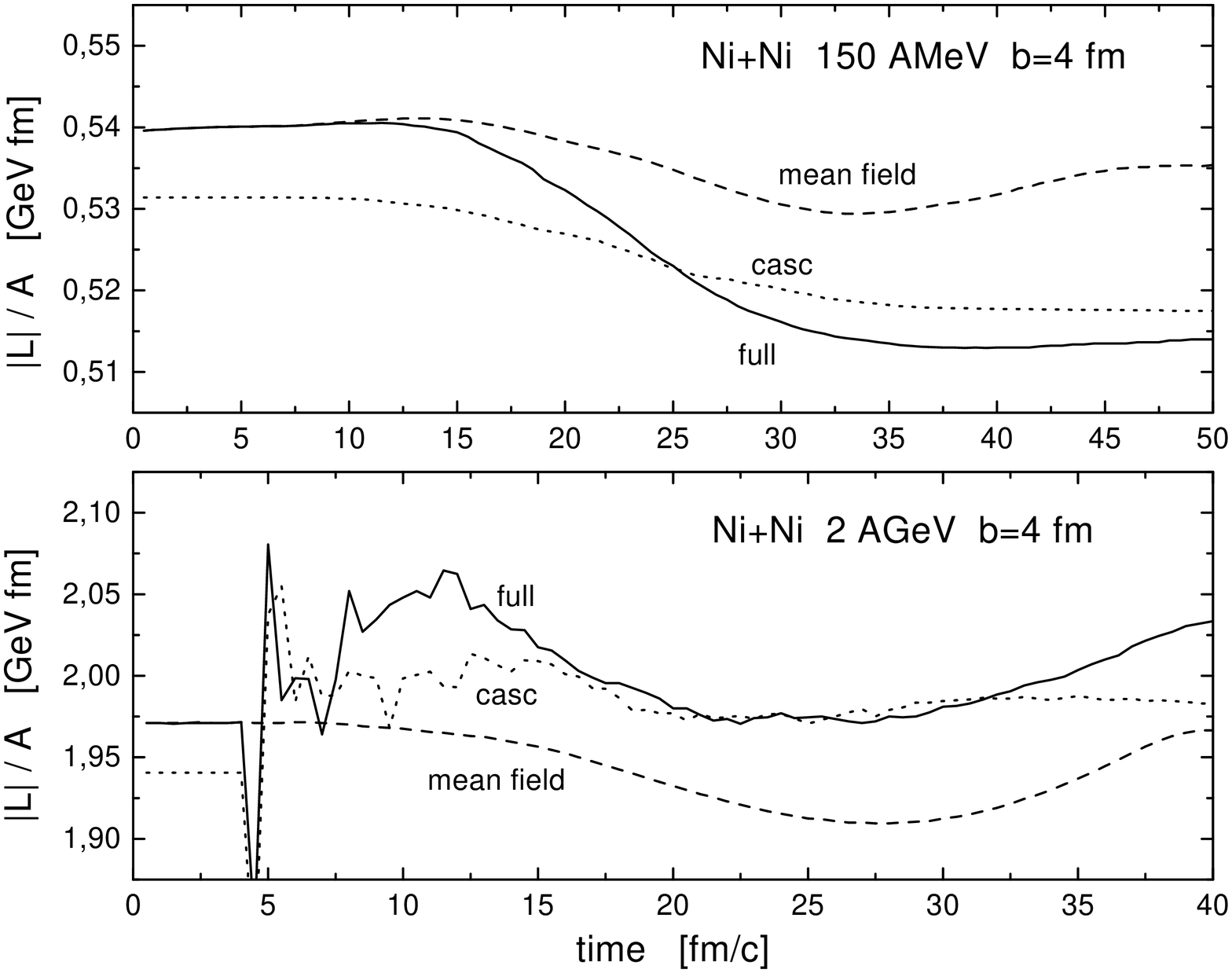}
\end{center}
\vspace{1cm}
\caption{}
\label{L_t_Ni}
\end{figure}

\begin{figure}[H]
\begin{center}
\includegraphics[angle=90,totalheight=21cm]{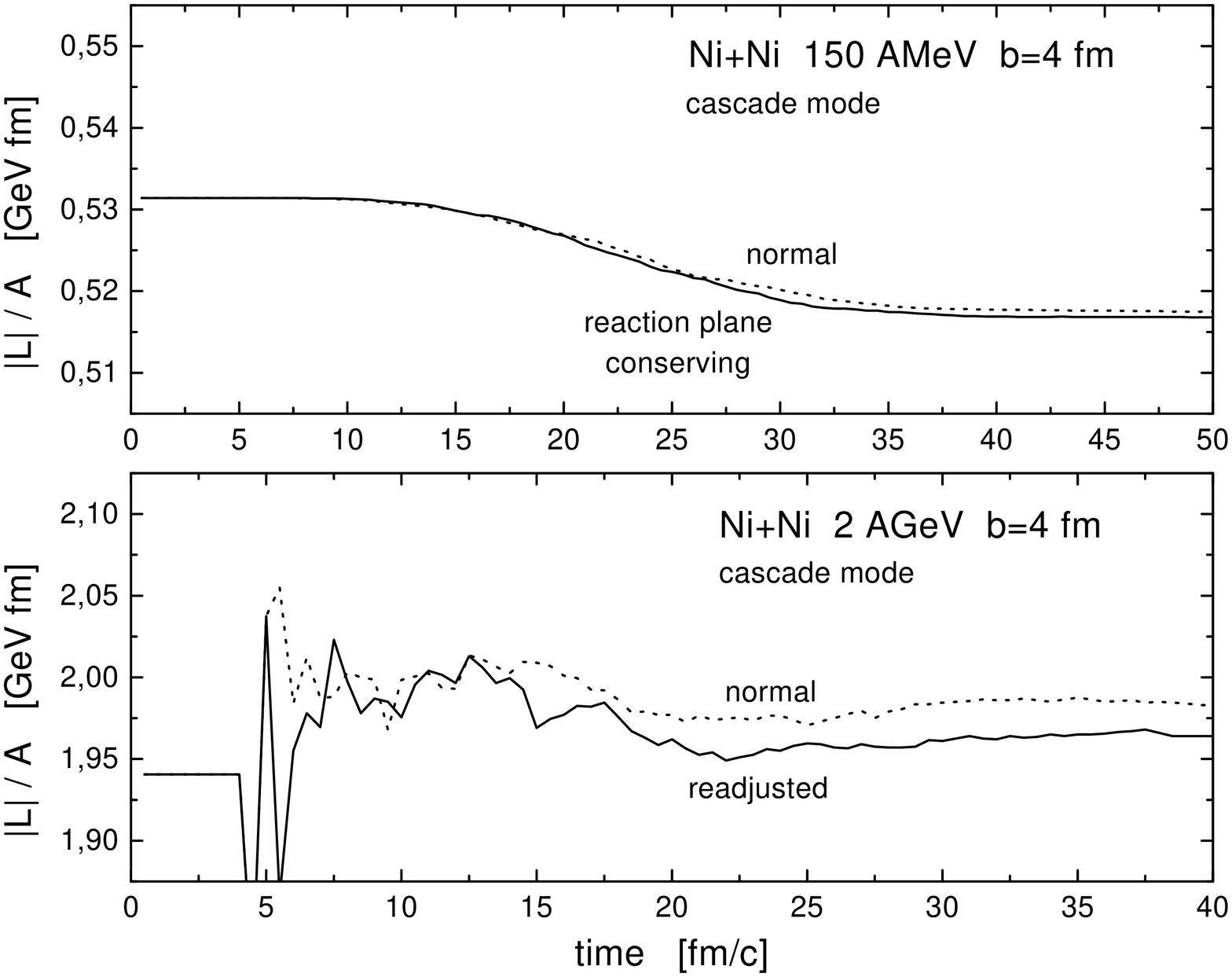}
\end{center}
\vspace{1cm}
\caption{}
\label{L_t_Ni_mod}
\end{figure}

\begin{figure}[H]
\begin{center}
\includegraphics[totalheight=20cm]{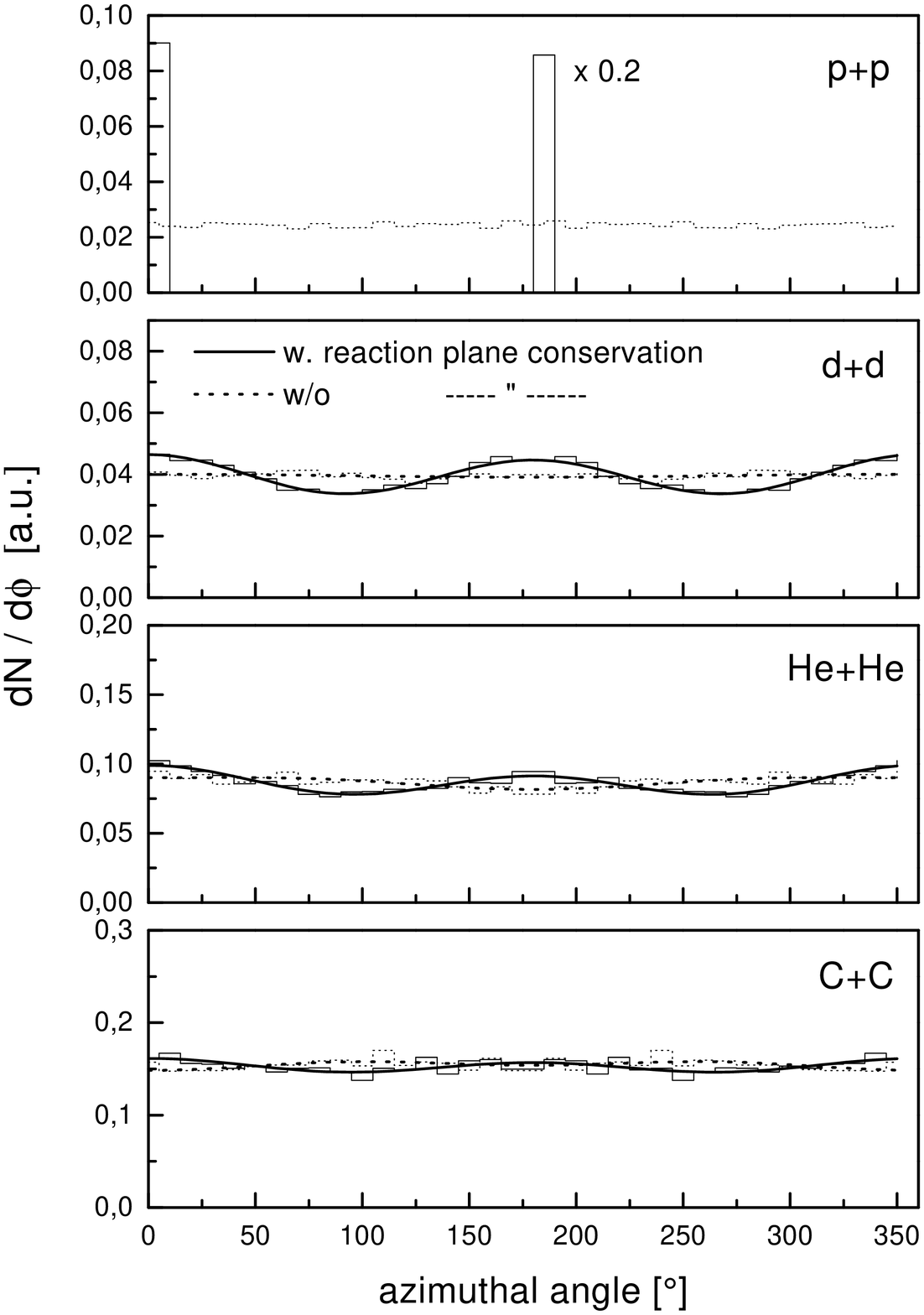}
\end{center}
\vspace{1cm}
\caption{}
\label{dNdphi}
\end{figure}

\begin{figure}[H]
\begin{center}
\includegraphics[angle=90,totalheight=22cm]{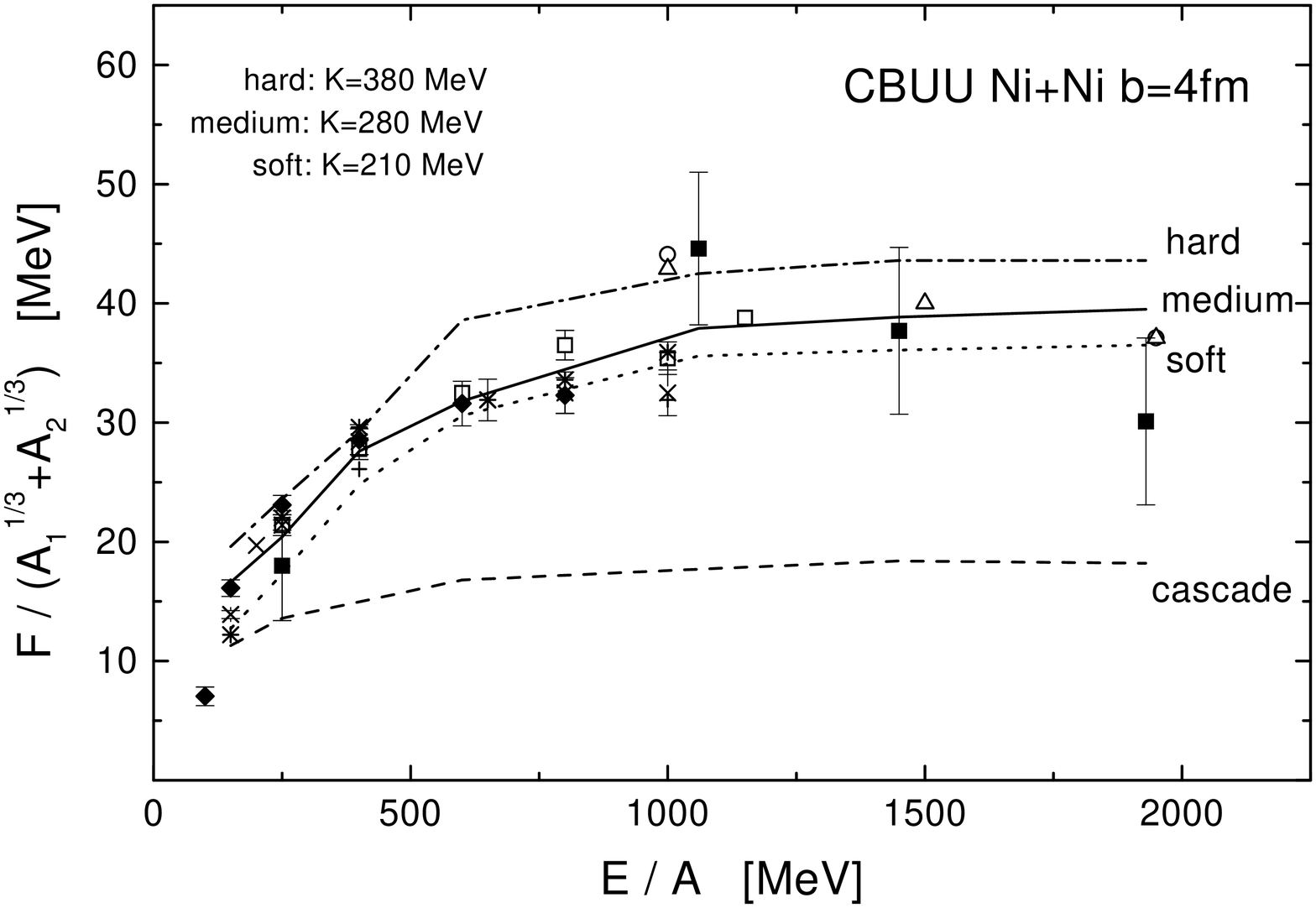}
\end{center}
\vspace{1cm}
\caption{}
\label{trfl_eos}
\end{figure}

\begin{figure}[H]
\begin{center}
\includegraphics[angle=90,totalheight=21cm]{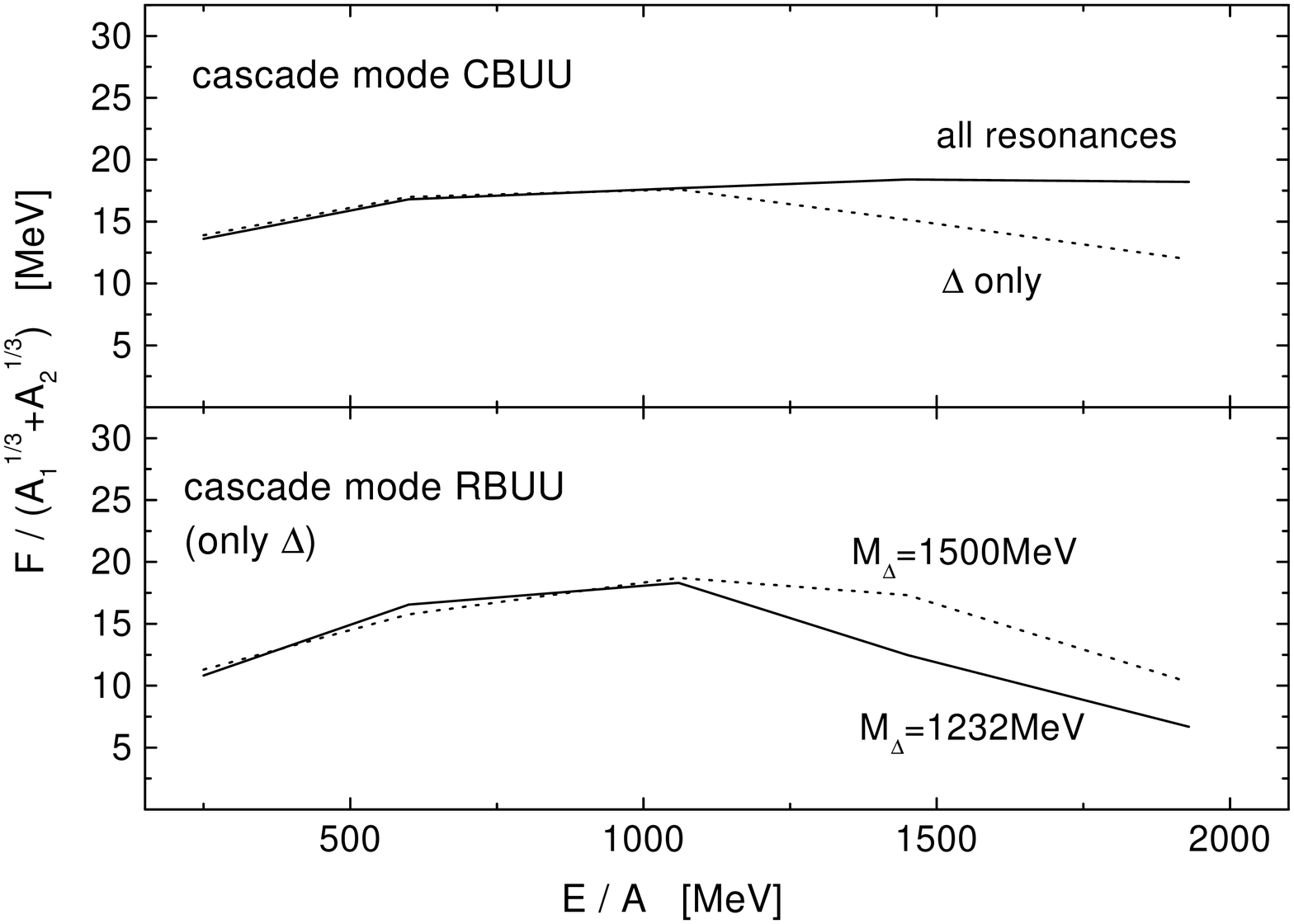}
\end{center}
\vspace{1cm}
\caption{}
\label{trfl_mres}
\end{figure}

%\begin{figure}[H]
%\begin{center}
%\includegraphics[totalheight=20cm]{hres_time.eps}
%\end{center}
%\vspace{1cm}
%\caption{}
%\label{hres_time}
%\end{figure}

%\begin{figure}[H]
%\centerline{
%\includegraphics[angle=0,totalheight=6.5cm]{hres1.eps}
%\mbox{}\hspace{1cm}\mbox{}
%\includegraphics[angle=0,totalheight=6.5cm]{hres2.eps}
%}
%\vspace{1.5cm}
%\centerline{
%\includegraphics[angle=-90,totalheight=6.5cm]{hres4.eps}
%\includegraphics[angle=-90,totalheight=6.5cm]{hres5.eps}
%}
%\vspace{7cm}
%\caption{}
%\label{hres}
%\end{figure}

%\begin{figure}[H]
%\begin{center}
%\includegraphics[totalheight=20cm]{hres_coll.eps}
%\end{center}
%\vspace{2cm}
%\caption{}
%\label{hres_coll}
%\end{figure}

%\begin{figure}[H]
%\vspace{2cm}
%\begin{center}
%\includegraphics[totalheight=10cm]{hres6.eps}\\
%\end{center}
%\vspace{7cm}
%\caption{}
%\label{hres6}
%\end{figure}

%\begin{figure}[H]
%\begin{center}
%\includegraphics[angle=90,totalheight=22cm]{trfl_hsd.eps}\\
%\end{center}
%\vspace{1cm}
%\caption{}
%\label{trfl_martin}
%\end{figure}

\begin{figure}[H]
\begin{center}
\includegraphics[totalheight=10cm]{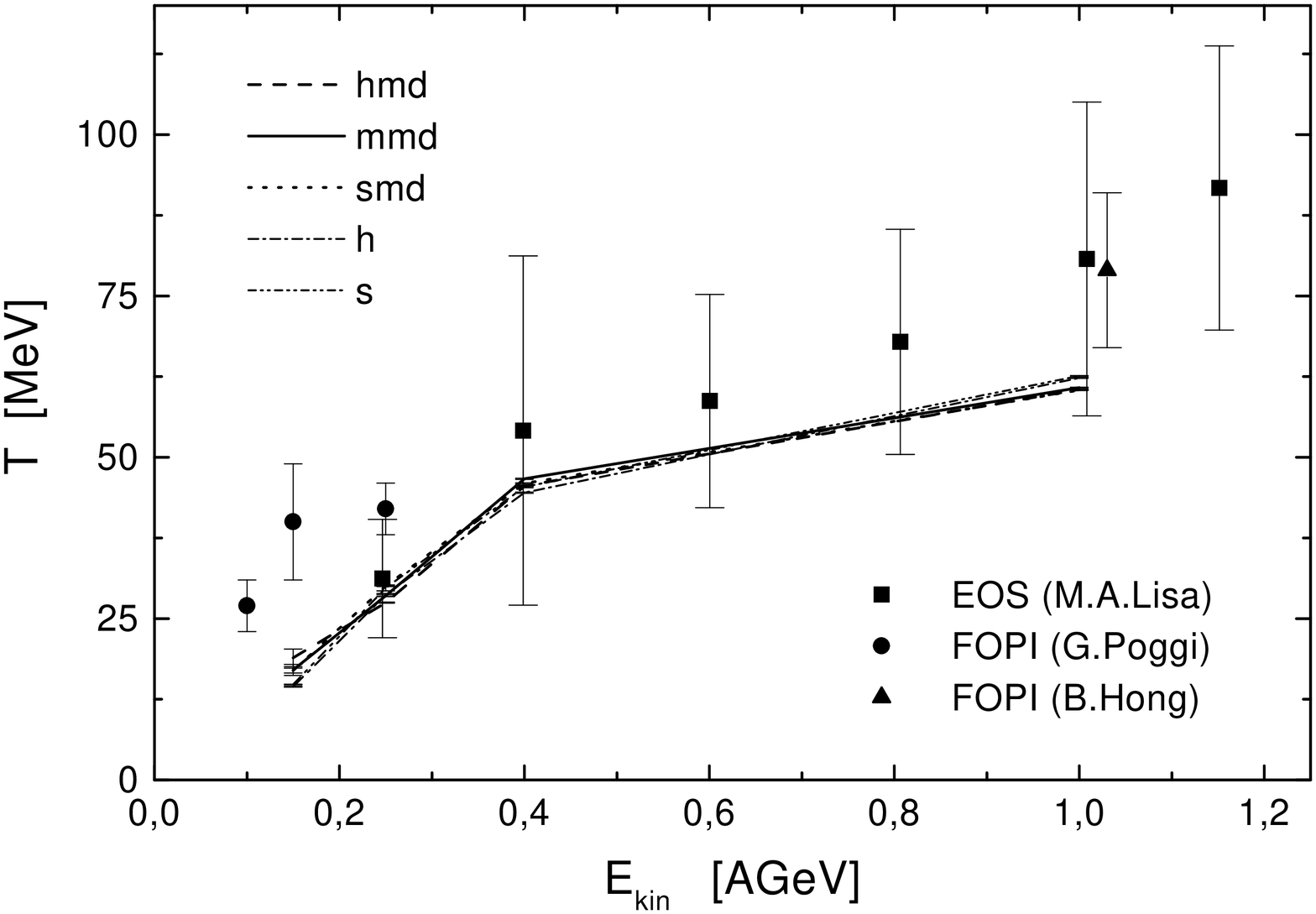}\\
\includegraphics[totalheight=10cm]{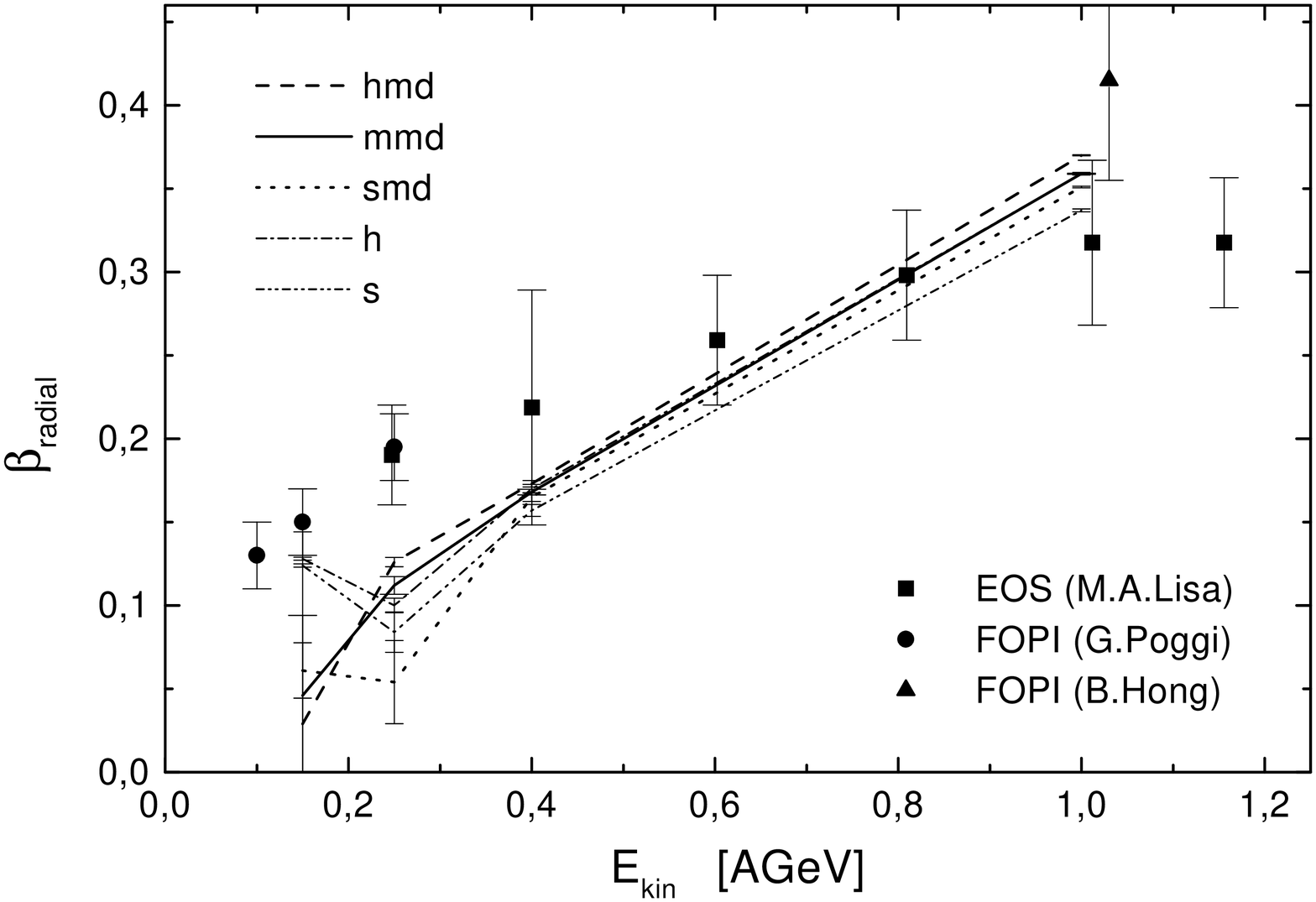}
\end{center}
\vspace{2cm}
\caption{}
\label{rad}
\end{figure}

\begin{figure}[H]
\begin{center}
\includegraphics[angle=90,totalheight=22cm]{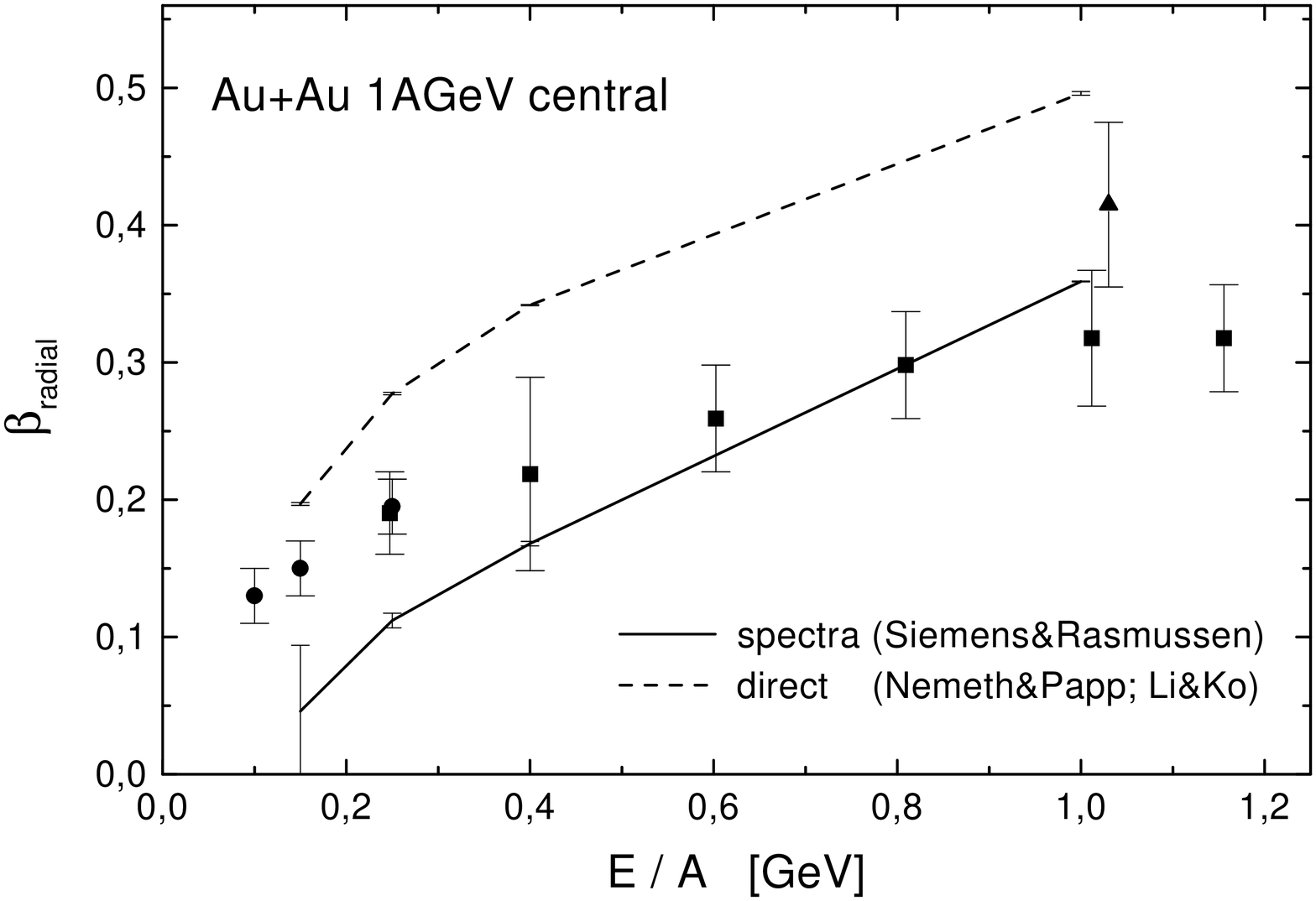}
\end{center}
\vspace{1cm}
\caption{}
\label{rad_b_SRNP}
\end{figure}

\begin{figure}[H]
\begin{center}
\includegraphics[totalheight=10cm]{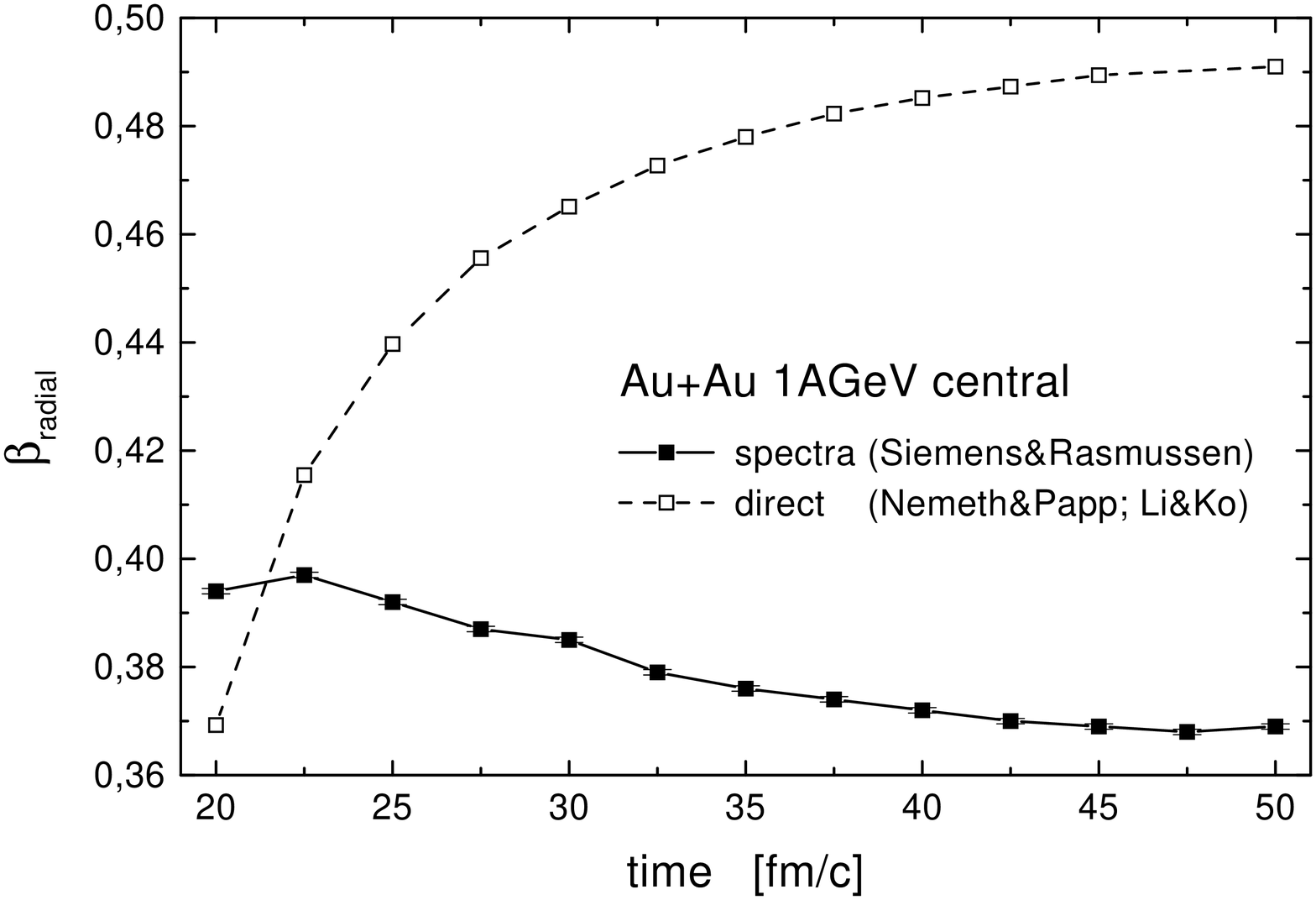}\\
\includegraphics[totalheight=10cm]{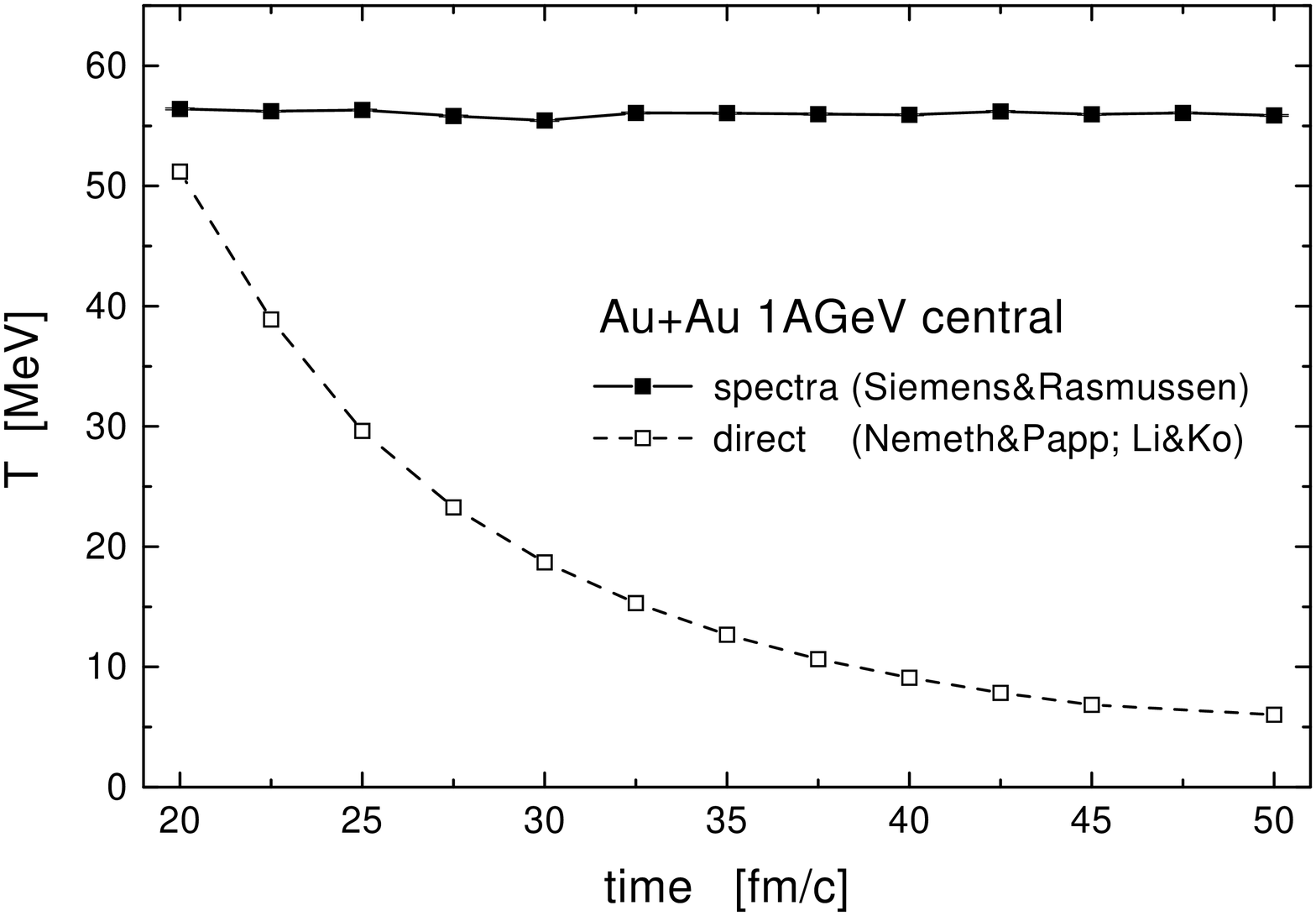}
\end{center}
\vspace{2cm}
\caption{}
\label{rad_t_SRNP}
\end{figure}

\begin{figure}[H]
\begin{center}
\includegraphics[angle=90,totalheight=22cm]{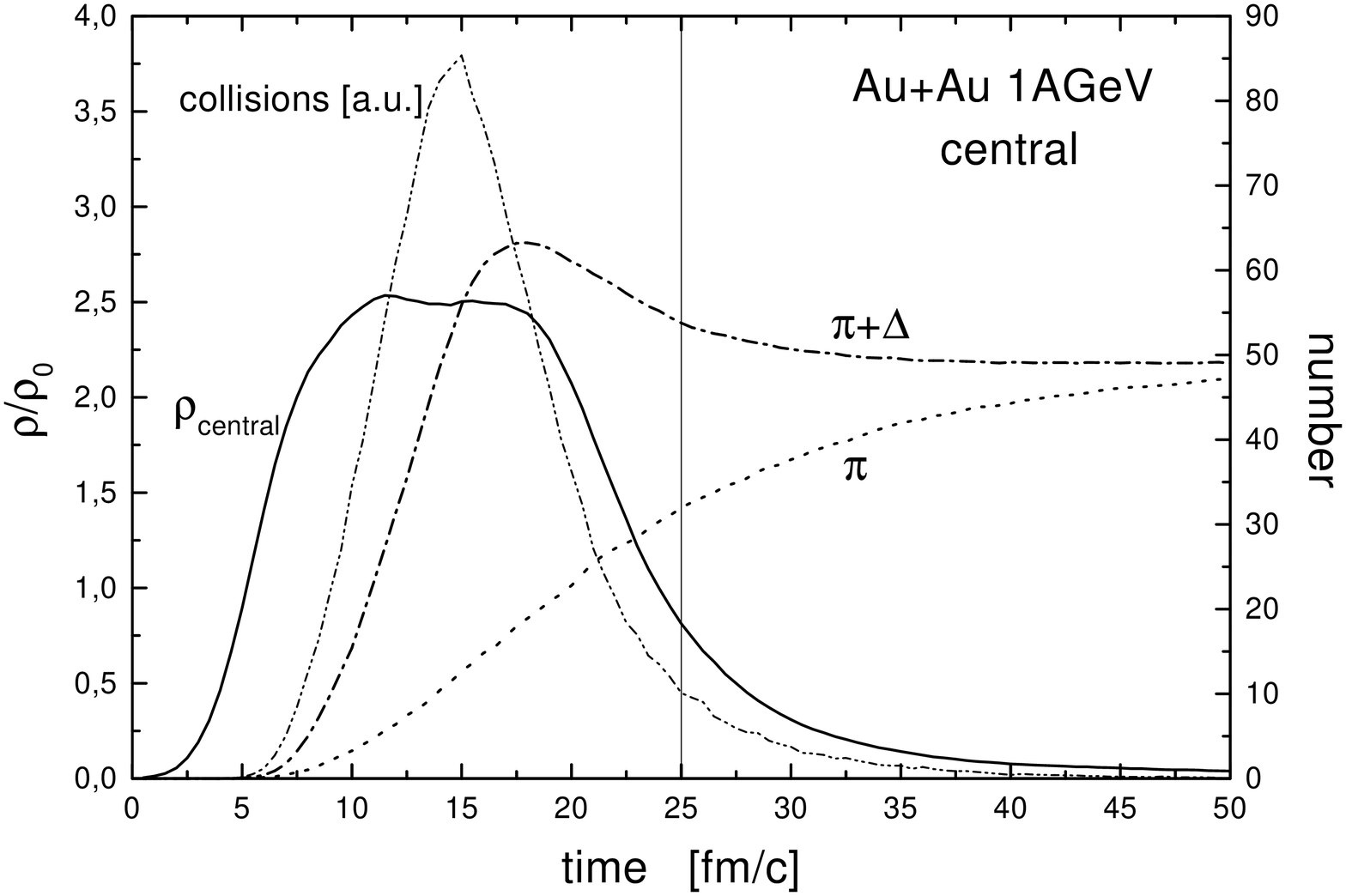}
\end{center}
\vspace{1cm}
\caption{}
\label{gyuri}
\end{figure}

\begin{figure}[H]
\begin{center}
\includegraphics[angle=90,totalheight=22cm]{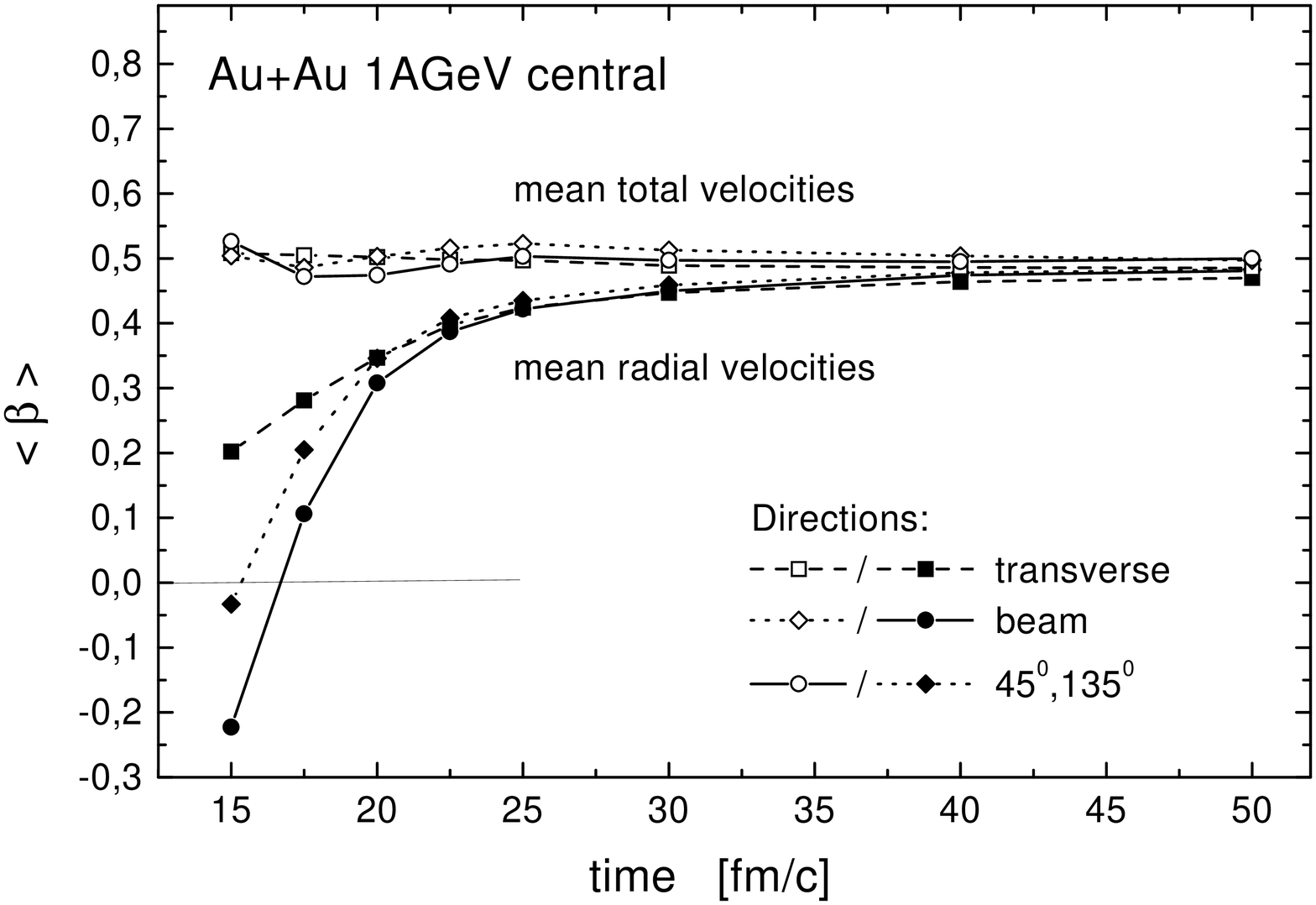}
\end{center}
\vspace{1cm}
\caption{}
\label{allbeta}
\end{figure}

\begin{figure}[H]
\begin{center}
\includegraphics[angle=90,totalheight=22cm]{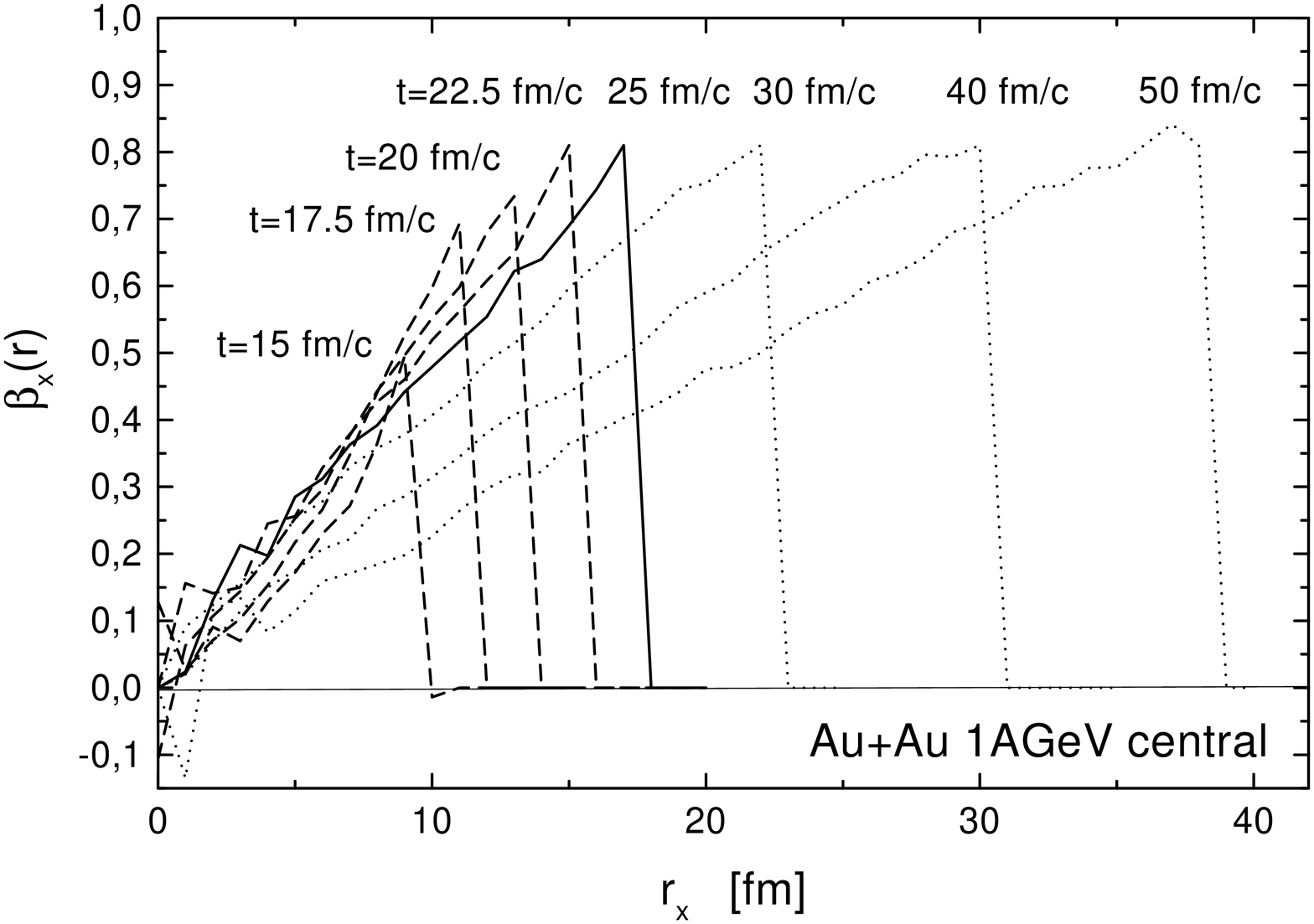}
\end{center}
\vspace{1cm}
\caption{}
\label{beta_x_t}
\end{figure}

\begin{figure}[H]
\begin{center}
\includegraphics[angle=90,totalheight=22cm]{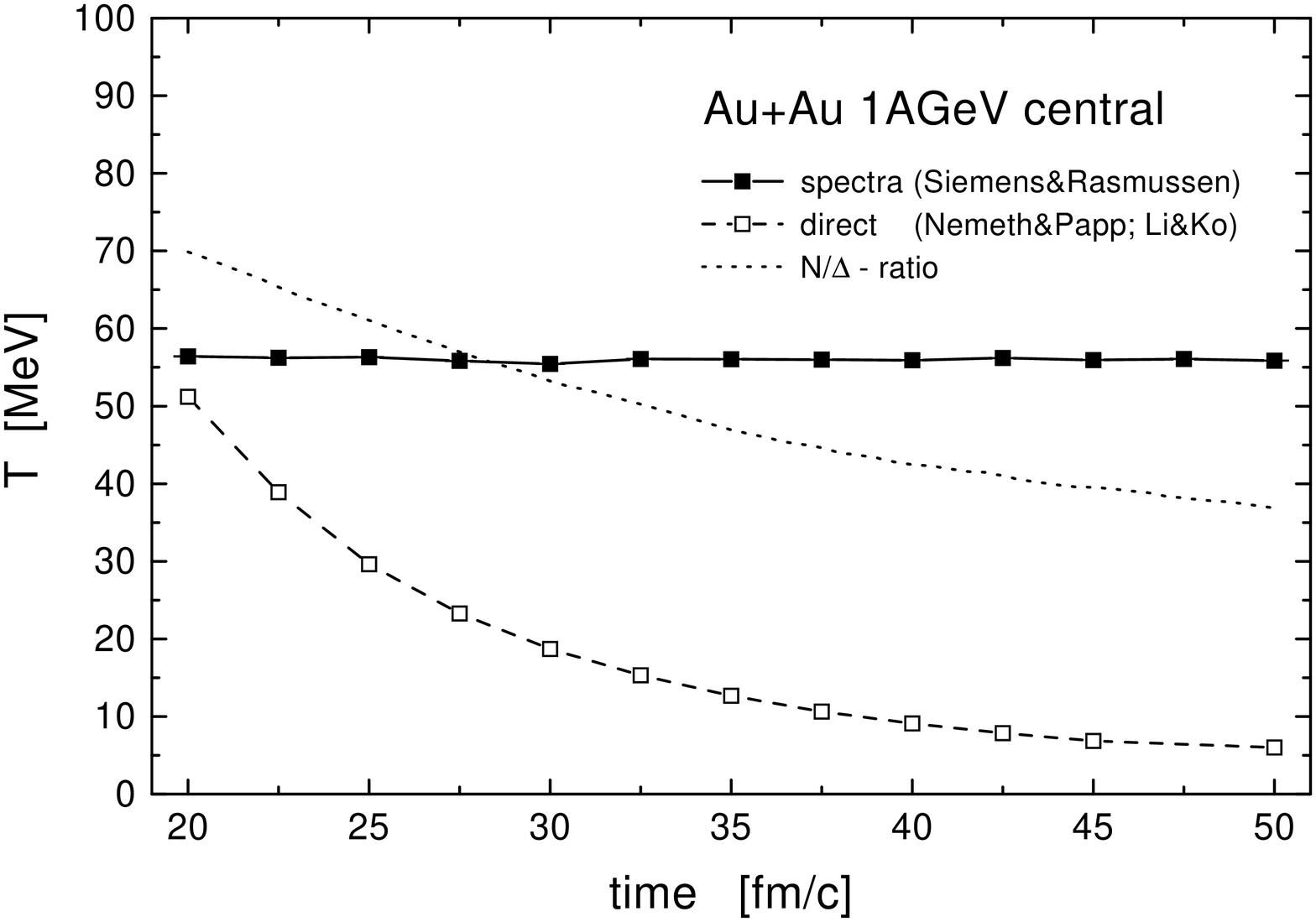}
\end{center}
\vspace{1cm}
\caption{}
\label{T_t_SRNPND}
\end{figure}

\begin{figure}[H]
\begin{center}
\includegraphics[totalheight=10cm]{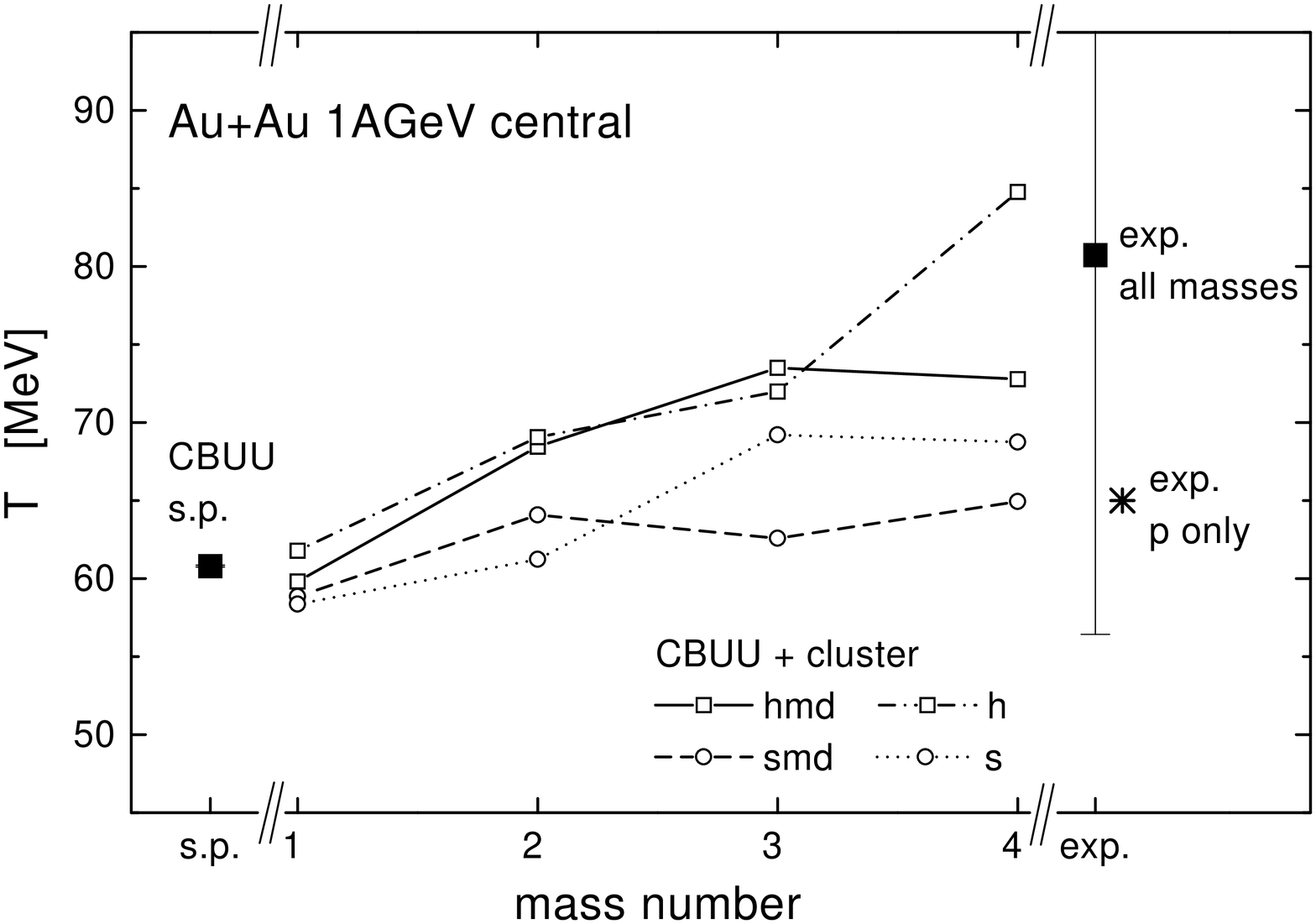}\\
\includegraphics[totalheight=10cm]{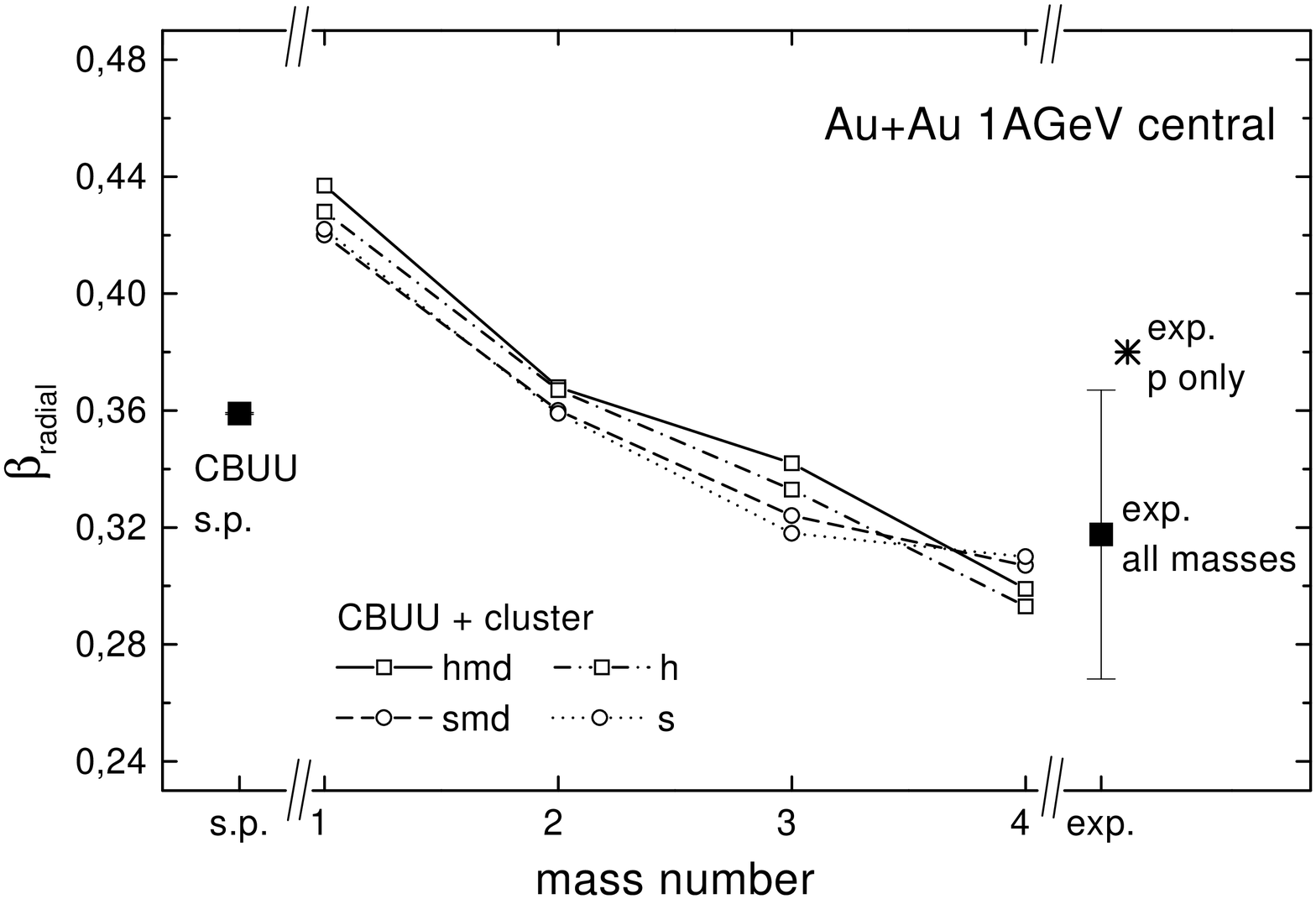}
\end{center}
\vspace{2cm}
\caption{}
\label{massdep_radflow}
\end{figure}

\begin{figure}[H]
\begin{center}
\includegraphics[angle=90,totalheight=22cm]{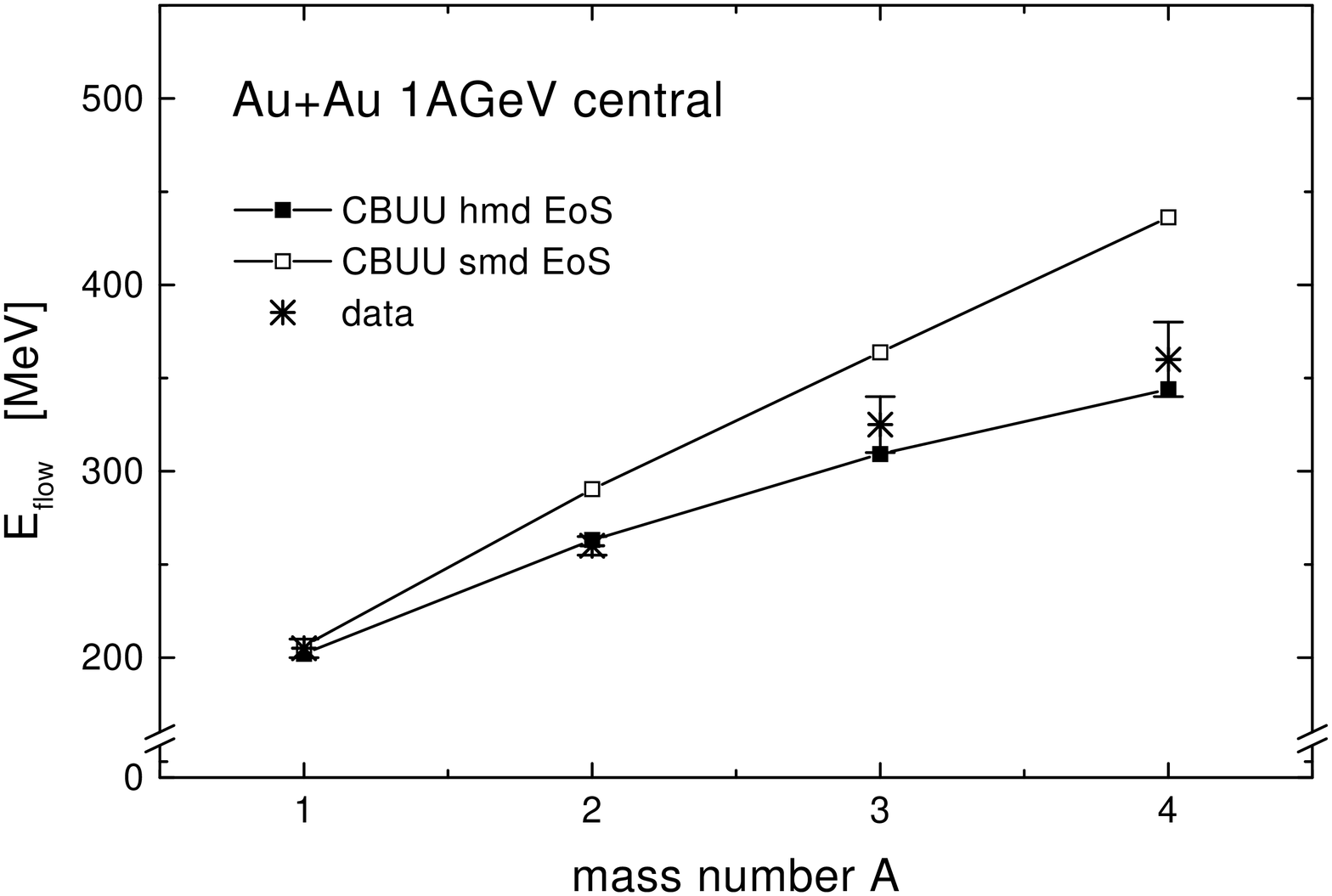}
\end{center}
\vspace{1cm}
\caption{}
\label{cl2_Eflow}
\end{figure}

\begin{figure}[H]
\begin{center}
\includegraphics[angle=90,totalheight=22cm]{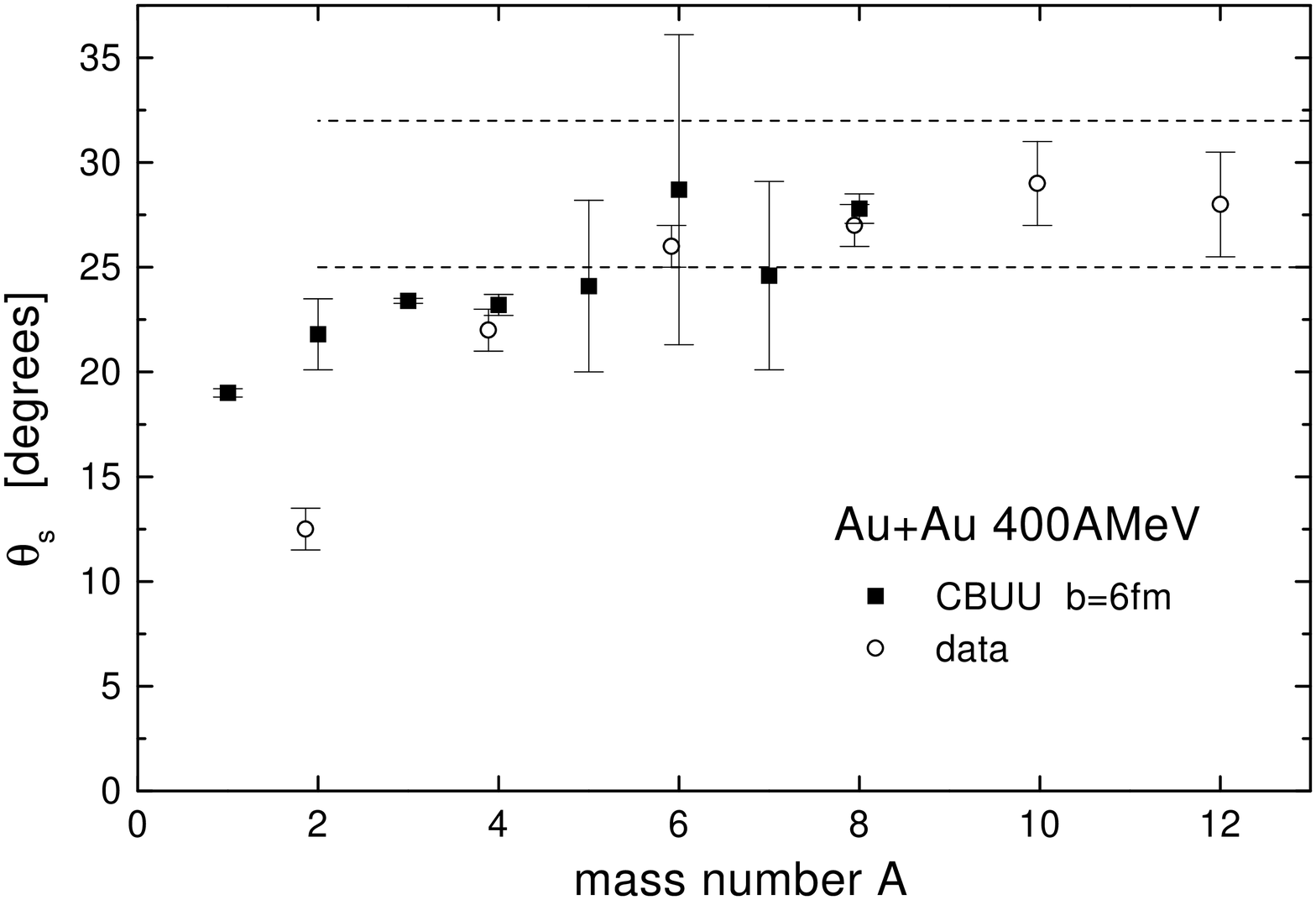}
\end{center}
\vspace{1cm}
\caption{}
\label{theta_s}
\end{figure}

\end{section}